\documentclass[prd,aps,preprint,showpacs,preprintnumbers,nofootinbib]{revtex4}
\usepackage{epsfig,footnote}
\usepackage{ulem}
\usepackage{color}
\usepackage{array}
\usepackage{amssymb}
\usepackage{amsmath}
\usepackage{graphicx,subfigure}
\usepackage{longtable}
\usepackage{verbatim}
\usepackage{amsfonts}
\usepackage{hyperref}
\usepackage{cancel}
\begin{document}
\title{Analysis of Charmless Two-body B decays in Factorization Assisted Topological Amplitude Approach}
\author{Si-Hong Zhou$^a$, Qi-An Zhang$^a$, Wei-Ran Lyu$^b$ and Cai-Dian L\"u$^a$ }
\affiliation{$^a$ Institute of High Energy Physics, YuQuanLu 19B, Beijing 100049, China;\\School of Physics, University of Chinese Academy of Sciences, YuQuanLu 19A, Beijing 100049, China;}
\affiliation{$^b$ Physics Department, Renmin University of China, ZhongGuanCun St. 59, Beijing 100872,  China}
\date{\today}
\begin{abstract}
We analyze charmless two-body non-leptonic B decays $B \to PP, PV$ under the framework
of factorization assisted topological amplitude approach, where $P(V)$ denotes a light
pseudoscalar (vector) meson. Compared with the conventional flavor diagram approach,
we consider flavor $SU(3)$ breaking effect assisted by factorization hypothesis
for topological diagram amplitudes of different decay modes, factorizing out the
corresponding decay constants and form factors.
The non-perturbative parameters of topology diagram magnitudes $\chi$ and strong phase $\phi$
are universal that can be extracted by $\chi^2$ fit from
current abundant experimental data of charmless B decays. The number of free parameters and
the $\chi^2$ per degree of freedom are both reduced comparing with previous analysis. With these
best fitted parameters, we predict branching fractions and $CP$ asymmetry parameters of
nearly 100 $B_{u,d}$ and $B_s$ decay modes. The long-standing
$\pi \pi$ and $\pi K$-$CP$ puzzles are resolved simultaneoulsy.
\end{abstract}
\maketitle

\section{Introduction}

Charmless two-body non-leptonic B decays are of importance for testing the standard model(SM).
They can be used to study CP violation via the interference of tree and penguin contributions.
They are also sensitive to signals of new physics that would change the small loop effects from
penguin diagrams. With regards to them, the BarBar and Bell experiments at the
$\mathrm{e}^{+}\mathrm{e}^{-}$ $B$-factories\cite{Bevan:2014iga} and LHCb experiment
at the Large Hadron Collider(LHC)\cite{Bediaga:2012py} have made great efforts in
studying B decays information in the past decades. Numerous data of branching fractions and
CP asymmetries of $B\to PP, PV$ decays, where $P(V)$ denotes a light
pseudoscalar (vector) meson, have been measured. In particular,
running at higher sensitivities and statistics, several $B_{s}$ decay channels have
been observed in LHCb experiment.
Such abundant experimental data have made it  possible to extract non-perturbative
parameters of hadronic decay amplitudes and to test theoretical calculations of
$B\to PP,PV$ decays.

In the theoretical side, as the non-leptonic B decays include hadronic decay amplitudes,
it requires complicated study of non-perturbative strong QCD dynamics. Furthermore, the charmless
$B$ decays not only involve tree topologies but also have
penguin loop diagrams that made the theoretical calculations more complex.
The measured large direct CP violation in charmless $B$ decays
indicates  the existence of  large strong phases, which mainly come  from non-perturbative QCD dynamics.
In the heavy b quark mass limit, we can factorize the perturbative calculable part from the non-perturbative hadronic matrix elements.
 The naive factorization approach \cite{Wirbel:1985ji} was first invented to
estimate the hadronic decay amplitudes, where they were factorized into
the product of perturbative hard kernels (local four quark operators) and non-perturbative objects
such as B to light form factors and decay constants of light pseudoscalar/vector mesons.
Then it was later improved to the generalized factorization approach\cite{Ali:1997nh}.
Based on the leading order power expansion of $\Lambda_{QCD}/\mathrm{m}_b$, where $\Lambda_{QCD}$
represents the typical non-perturbative QCD hadronic scale, $\mathrm{m}_b$ is $b$ quark mass,  the
QCD factorization (QCDF) \cite{Beneke:2000ry}, the perturbative QCD (PQCD)\cite{Lu:2000em},
and the soft-collinear effective theory (SCET)\cite{Bauer:2000yr}   have been developed   recently.
Great theoretical progress have been made in these perturbative QCD approaches.
However, it is impossible to calculate to all order of power expansions, thus
some strong QCD dynamics information would be lost in these perturbative approaches.
With the very high precision  of experimental data,
  the leading-order theoretical calculation of $\Lambda_{QCD}/\mathrm{m}_b$ expansion  is not  enough. For example, QCDF \cite{Cheng:2009cn} need to include
a large penguin annihilation contribution (as free parameter) to enhance the
branching fractions and direct CP asymmetry of penguin-dominated charmless $B$ decays.
The same puzzle also appeared in SCET \cite{Bauer:2004tj}, where
the penguin annihilation contribution in QCDF is replaced by
the power suppressed (but with larger numerical contribution than the leading terms) nonperturbative charming penguin effect.
All these power corrections are not able to calculate perturbatively but
need to be fitted  from experiments.
There are also some experimental puzzles to be solved for those perturbative approaches. The perturbative calculation predict the same sign of direct CP asymmetry in $B^{\pm}\to \pi^{0} K^{\pm}$
and in $B^{0}\to \pi^{\mp} K^{\pm}$  decays, which is conflict with experimental data.
The calculated branching ratio of $ B^{0}\to \pi^{0}\pi^{0}$ in perturbative approaches
is several times smaller than the measured one. These long-standing puzzles
are sensitive to the non-factorizable color suppressed emission  diagram.
Although some soft and sub-leading effects were taken into account in QCDF \cite{Cheng:2009cn}
and PQCD \cite{Li:2005kt}, the $B \to \pi \pi$ puzzle was still left in
the conventional factorization theorem.  Recently,  an additional Glauber phase is introduced to solve this puzzle \cite{Li:2014haa}.

Unlike the above mentioned perturbative approaches, some model-independent approaches were
introduced to analyze the charmless $B$ decays, such as global $SU(3)/U(3)$ flavor symmetry analysis \cite{he} and
flavor topological diagram approach \cite{chiang,Cheng:2014rfa}.
They do not apply factorization in QCD, leaving all perturbative or non-perturbative QCD effects extracted from experimental data.
In \cite{he}, they related relevant decay amplitudes using $SU(3)/U(3)$ group decomposition and then extract them form experimental data.
For the flavor topological diagram approach, they group different contributions according to the electroweak topological diagram,
since electroweak interaction naturally factorize from QCD interaction. Each
topological diagram amplitude  including all strong interactions with strong phase are to be  extracted from  experimental  data.
However, in order to reduce the number of free parameters, it needs to apply the flavor $SU(3)$ symmetry to
relate  topological diagram parameters of different decay modes.
In fact, the flavor $SU(3)$ symmetry is broken. Nowadays, $SU(3)$ breaking effect have to be considered to compare the
theoretical results  with the  precise
experimental data.
It is also observed in the flavor topological diagram analysis that
there are large differences among the three types of $B\to PP$, $B\to PV$ and $B\to VP$ decays
due to different pseudoscalar and vector final states.
Therefore, they have to fit three different sets of parameters for
the three types of B decays respectively \cite{Cheng:2014rfa}.
There are  too many parameters to be fitted,
its prediction power was reduced.

In view of the above complexity and incompleteness in power correction
of factorization approaches and the limitation of the conventional flavor
topological diagram approach, a new method called
factorization-assisted-topological-amplitude (FAT) approach was proposed
in studying the two-body hadronic decays of D mesons \cite{Li:2012cfa,Li:2013xsa}.
Aiming  to include all non-factorizable/non-perturbative QCD contributions
compared to factorization approaches, it adopts the formalism of  flavor topological diagram approach.
However, different from the conventional
flavor topological diagram approach, it had included $SU(3)$ breaking effect
in each flavor topological diagram assisted by factorization hypothesis.
The FAT approach applied in D mesons decays \cite{Li:2012cfa,Li:2013xsa}
was in great success to resolve the long-standing puzzle from the large difference of
 $D^{0}\to \pi^{+}\pi^{-}$ and $D^{0}\to K^{+}K^{-}$ branching fractions, due to large $SU(3)$ breaking effects. It also predicted 0.1\% of direct CP asymmetry difference between these two decay channels, later confirmed by the LHCb experiment \cite{Qin:2014nxa}.
With an intermediate charm quark scale, the two-body charmed  meson decays of B meson also encounter large $SU(3)$ breaking effects \cite{Zhou:2015jba}.
With only 4 parameters fitted from 31 experimental observations,
we predict  120 decay modes, some of which are  tested by the available experimental data \cite{Zhou:2015jba}.

In this work, we will analyze the charmless $B \to PP$, $PV$ decays in the FAT approach.
Being different from the two-body charmed B meson decays
with only tree topologies, penguin topological diagrams
enhanced by CKM matrix elements will contribute to these charmless B meson decays.
These loop effect will be more complicated than the calculation to tree level diagrams.
 More theoretical parameters are needed to describe these penguin topological amplitudes and more experimental observables, such as CP asymmetry parameters.
Specifically, including penguin topological contributions,
we will fit 14 parameters from 37 experimental measured branching fractions and 11
CP asymmetry parameters of $B \to PP$ and $B \to PV$ decays.
The number of free parameters is significantly reduced from the previous topological diagram approach with much less $\chi^2$ per degree of freedom. The long-standing $B \to \pi \pi $ puzzle and $ B \to\pi K$ CP puzzle
are resolved consistently.

In Sec.\ref{FAT}, we parameterize the
tree and penguin topological amplitudes of charmless $B \to PP$, $PV$ decays
in  the FAT approach. The numerical results and discussions are presented
in Sec.\ref{results}. Sec.\ref{conclusion} is the conclusion.

\section{ Factorization of Decay Amplitudes for different Topological diagrams}\label{FAT}

  \begin{figure}
  \begin{center}
  \scalebox{1}{\epsfig{file=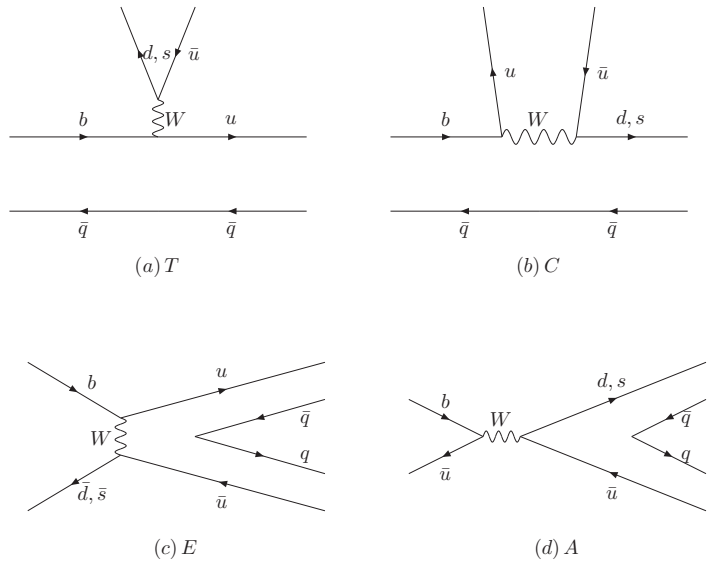}}
  \caption{Topological tree diagrams contributing to
      $B\to PP $ and $B\to PV$ decays:
  (a) the color-favored   tree emission diagram, $T$;
  (b) the color-suppressed  tree emission diagram, $C$;
  (c) the $W$-exchange     diagram, $E$ and
  (d) the $W$-annihilation     diagram, $A$.}
  \label{Tree}
  \end{center}
  \end{figure}

\begin{figure}[htbp]
  \begin{center}
  \scalebox{1}{\epsfig{file=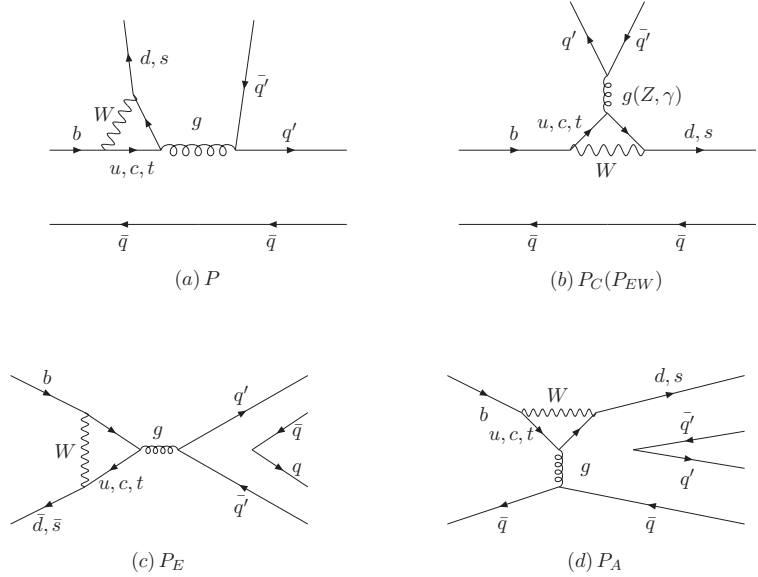}}
  \caption{Topological penguin diagrams contributing to
      $B\to PP $ and $B\to PV$ decays:
  (a) the color-favored  QCD-penguin diagram, $P$;
  (b) the flavor-singlet   QCD-penguin diagram, $P_C$
  and EW-penguin   diagram $P_{EW}$;
  (c) the exchange type QCD-penguin diagram, $P_E$ and
  (d) the    QCD-penguin annihilation diagram, $P_A$.}
 \label{Penguin}
  \end{center}
  \end{figure}

The two body charmless B decays are flavor changing weak decays. They are induced by the quark level diagrams classified by leading order (tree diagram) and 1-loop level (penguin diagram) weak interactions.
For different B decay final states, the tree level weak decay diagram can contribute via different orientations: the so-called  color-favored tree emission diagram $T$,
color-suppressed tree emission diagram $C$,
$W$-exchange tree diagrams $E$ and
$W$ annihilation tree diagrams $A$, which are shown in Fig.\ref{Tree}, respectively. These tree level diagrams have already been studied in the previous D meson decays  \cite{Li:2012cfa,Li:2013xsa} and charmed meson final state B decays \cite{Zhou:2015jba}.
Similarly, the 1-loop penguin diagram can also be classified as   5-types:
color-favored QCD penguin emission diagram $P$,
color-suppressed QCD penguin emission diagram $P_C$,
$W$-annihilation penguin diagram $P_A$,
the $W$ penguin exchange diagram $P_E$ and
electro-weak penguin emission diagram $P_{EW}$, shown in Fig.\ref{Penguin}.

In the QCD factorization approaches, one try to calculate the QCD corrections to the specific weak diagrams or effective four quark operators order by order. The decay amplitude for each decay is calculated in the factorization framework by the heavy quark expansion. In this work, to avoid the dependence of specific factorization approach, we extract    the two-body hadronic weak
decay amplitude of different topological diagram  from the experimental data by the $\chi^2$ fit.
Therefore all strong interaction effects,
the factorization and non-factorization contributions, perturbative and non-perturbative QCD corrections
are all determined by experimental measurements. This is the idea of conventional topological diagram approach \cite{Cheng:2014rfa}. In order to have predictive power, one has to assume the flavor $SU(3)$ symmetry, reducing the number of independent parameters. The precision of this topological diagram approach is then limited to the order of $SU(3)$ breaking.  In the FAT approach, we will try to recover the $SU(3)$ breaking effects, further reducing the number of free parameters by fitting all the decay channels.

Let's start from tree amplitudes shown in Fig.\ref{Tree}. In the conventional topological diagram approach, the color favored tree amplitude (T) is tuned to be a real number, with 6 parameters (magnitudes and phases) for three other amplitudes. However, these 7 parameters have to be tripled for $B\to PP$, $B\to PV$ and $B \to VP$ decays, since there is a non-negligible difference between pseudo-scalar and vector mesons. In this work, we will try to parametrize these three kinds of decays together.
The color-favored $T$ topology shown in Fig.\ref{Tree}(a) is
proved factorization to all orders of $\alpha_{s}$ expansion in QCD factorization, perturbative QCD, and soft-collinear-effective theory. Their numerical results also agree to each other in different approaches. Thus, to reduce one free parameter,  we will just use their theoretical results, not fitting from the experiments:
\begin{align}\label{eq:T}
T^{P_{1}P_{2}}&=i\frac{G_{F}}{\sqrt{2}}V_{ub}V_{uq^{'}}a_{1}
(\mu)f_{p_{2}}(m_{B}^{2}-m_{p_{1}}^{2})F_{0}^{BP_{1}}(m_{p_{2}}^{2}),\\
T^{PV}&=\sqrt{2}G_{F}V_{ub}V_{uq^{'}}a_{1}
(\mu)f_{V} m_{V}F_{1}^{B-P}(m_{V}^{2})(\varepsilon^{*}_{V}\cdot p_{B}),\\
T^{VP}&=\sqrt{2}G_{F}V_{ub}V_{uq^{'}}a_{1}
(\mu)f_{P} m_{V}A_{0}^{B-V}(m_{P}^{2})(\varepsilon^{*}_{V}\cdot p_{B}),
\end{align}
where the superscript  of $T^{P_{1}P_{2}}$ denote the final mesons
are two pseudoscalar mesons, $T^{PV(VP)}$ for recoiling mesons are
pseudoscalar meson (vector meson).
 $a_1(\mu)$ is the effective Wilson coefficient from short distance QCD corrections, where the
factorization scale $\mu$ is insensitive to different  final state mesons.
 Therefore we can choose it within a certain range arbitrarily and
set it at the point $\mu=\mathrm{m_b}/2=2.1 \mathrm{GeV}$.
$a_1(\mu)$ at this scale is $1.05$.
$f_{P_{2}}$($f_{P}$) and $f_{V}$ are the decay constants of emissive pseudoscalar meson
and vector meson, respectively.   $F_{0}^{BP_{1}}$ ($F_{1}^{B-P}$) and $A_{0}^{B-V}$ are the
  form factors of $B\to P$ and $B\to V$ transitions, respectively.
$\varepsilon^{*}_{V}$ is the polarization vector of vector meson and
$p_{B}$ is the 4-momentum of $B$ meson.

For the color suppressed $C$ topology, dominated by non-factorization contributions,
it is least-understood by us although having been calculated up
to next-to-leading order in the factorization methods.
The next-to-leading order corrections in factorization framework could not resolve the $\pi \pi$ and $\pi K$ puzzles
strongly sensitive to this $C$ topology contribution.
A large $C$ contribution with large strong phase (mostly non-perturbative) can resolve the so called  $\pi K$ puzzle. However, it is not possible to explain  the  $\pi \pi$ puzzle: theoretically $Br(B^0\to \pi^0 \pi^0) < Br(B^0\to \pi^0 \rho^0) < Br(B^0 \to \rho^0 \rho^0) $, but experimentally it is in the inverse order.
 In the conventional topological diagram approach   \cite{Cheng:2014rfa}, the authors introduced two parameters (amplitude and phase) in the $B\to PP$ modes and another four parameters in the $B\to PV$, $VP$ modes for this diagram to be fitted from experimental data.
To our knowledge, this inverse order can be understood only in the formalism of Glauber gluons introduced in ref.\cite{Li:2014haa}, where extra phase was introduced for the pseudo-scalar meson (Goldstone boson) emission diagram.
Inspired by these studies, We parameterize the $C$ diagram magnitude and associate phase as
$\chi^{C}$ and $\mathrm{e}^{i\phi^{C}}$ in $B\to PP$, $VP$ decays and
$\chi^{C^{\prime}}\mathrm{e}^{i\phi^{C^{\prime}}}$ in $B\to PV$, respectively
to distinguish cases in which the emissive meson is pseudo-scalar or vector:
\begin{align}\label{eq:C}
C^{P_{1}P_{2}}&=i\frac{G_{F}}{\sqrt{2}}V_{ub}V_{uq^{'}}\chi^{C}\mathrm{e}^{i\phi^{C}}
        f_{p_{2}}(m_{B}^{2}-m_{p_{1}}^{2})F_{0}^{BP_{1}}(m_{p_{2}}^{2}),\nonumber\\
C^{PV}&=\sqrt{2}G_{F}V_{ub}V_{uq^{'}}\chi^{C^{\prime}}\mathrm{e}^{i\phi^{C^{\prime}}}
         f_{V} m_{V}F_{1}^{B-P}(m_{V}^{2})(\varepsilon^{*}_{V}\cdot p_{B}),
\nonumber\\
C^{VP}&=\sqrt{2}G_{F}V_{ub}V_{uq^{'}}\chi^{C}\mathrm{e}^{i\phi^{C}}
        f_{P} m_{V}A_{0}^{B-V}(m_{P}^{2})(\varepsilon^{*}_{V}\cdot p_{B}),
\end{align}
where the decay constants and form factors $f_{P}$, $f_{V}$,$F_{0}^{BP_{1}}$, $F_{1}^{B-P}$ and $A_{0}^{B-V}$ characterizing the $SU(3)$ breaking effects
are factorized out.

The W-exchange $E$ topology is  non-factorization in QCD factorization approach. It is expected smaller
than emission diagram     due to helicity suppression.   We   use $\chi^{E}$, $\mathrm{e}^{i\phi^{E}}$ to represent
the magnitude and its strong phase for all decay modes:
\begin{align}\label{eq:E}
E^{P_{1}P_{2}} &=i\frac{G_{F}}{\sqrt{2}}V_{ub}V_{uq^{'}} \chi^{E} \mathrm{e}^{i\phi^{E}}
f_{B}m_{B}^{2}(\frac{f_{p_{1}}f_{p_{2}}}{f_{\pi}^{2}}),\nonumber\\
E^{PV,VP} &=\sqrt{2}G_{F}V_{ub}V_{uq^{'}}\chi^{E} \mathrm{e}^{i\phi^{E}}
(\mu)f_{B}m_{V}(\frac{f_{P}f_{V}}{f_{\pi}^{2}})(\varepsilon^{*}_{V}\cdot p_{B}),
\end{align}
Considering flavor $SU(3)$ breaking effects, we multiply decay constants of
three mesons $f_{B}$,$f_{p_{1}}(f_P)$ and $f_{p_{2}}(f_V)$ in each amplitude.
In order to make parameters $\chi^{E}$ and $\mathrm{e}^{i\phi^{E}}$ dimensionless,
a normalization factor ${f_{\pi}^{2}}$ is introduced. Actually, dimensionless parameters
$\chi^{E}$, $\mathrm{e}^{i\phi^{E}}$ are defined from $B \to \pi \pi$ decays.
  Other processes are related by different
decay constant factors $\frac{f_{p_2} f_{p_1}(f_P f_V)}{f_{\pi}f_{\pi}}$.
The last diagram in Fig.\ref{Tree}(d) is the so called W-annihilation topology. As discussed in ref.\cite{Cheng:2014rfa}, its contribution is negligible. We will also ignore it in this paper.

The penguin topological diagrams are grouped into QCD penguin and
electro-weak penguin (EW penguin) topologies. In terms of QCD penguin
diagram amplitude, we consider all contributions from every topological
diagram in Fig.\ref{Penguin}, where topology $P$ contributes most.
The leading contribution from topology $P$ diagram is similar to the color favored tree diagram T, which is proved factorization in various QCD-inspired approaches, such as QCD factorization \cite{Cheng:2009cn}, perturbative QCD \cite{Lu:2000em} and soft-collinear effective theory \cite{Wang:2008rk}. They give very similar numerical results proportional to the Wilson coefficient $a_{4}$, related to the
 QCD penguin operators $O_{3},O_{4}$. Therefore, in the same spirit of T diagram, we will not fit this contribution from the experimental data, but predict its contribution from QCD calculations for all the three type of  $B \to PP$, $B\to VP$ and $B\to PV$ decays. This is not the whole story. All these approaches predict large extra contribution in this topology related to the  effective four-quark operators $O_{5},~O_{6}$, which is also called the ``chiral enhanced'' penguin contributions. Since this chiral enhancement  only contributes to the pseudo-scalar meson (Goldstone boson) emission diagram, we will include it only in $B \to PP$  and $B\to VP$ decays,
which can be parameterize as $r_{\chi}\chi^{P}$, $\mathrm{e}^{i\phi^{P}}$ in Eq.(\ref{eq:P})
with $r_{\chi}$   the chiral factor of pseudo-scalar meson.  The decay amplitude for the penguin diagram $P$ is then parameterized with only two free parameters for all the three categories of $B \to PP$, $B\to VP$ and $B\to PV$ decays, as
\begin{align}\label{eq:P}
P^{PP}&=-i\frac{G_{F}}{\sqrt{2}}V_{tb}V_{tq^{'}}^{*}\left[a_{4}(\mu)+\chi^{P}\mathrm{e}^{i\phi^{P}}r_{\chi}\right]
f_{p_{2}}(m_{B}^{2}-m_{p_{1}}^{2})F_{0}^{BP_{1}}(m_{p_{2}}^{2}),\nonumber
\\
P^{PV}&=-\sqrt{2}G_{F} V_{tb}V_{tq^{'}}^{*}a_{4}(\mu) f_{V}m_{V}F_{1}^{B-P}m_{V}^{2}
(\varepsilon^{*}_{V}\cdot p_{B}),\nonumber\\
P^{VP}&=-\sqrt{2}G_{F}V_{tb}V_{tq^{'}}^{*}\left[a_{4}(\mu)-\chi^{P} \mathrm{e}^{i\phi^{P}}r_{\chi}\right]
 f_{P}m_{V}A_{0}^{B-V}(m_{P}^{2})(\varepsilon^{*}_{V}\cdot p_{B}).
\end{align}

The so called penguin annihilation diagram $P_A$ shown in Fig.\ref{Penguin}(d) was considered as a power correction to $P$, calculated perturbatively
in PQCD approach \cite{Lu:2000em}, parameterized   as $\rho_{A}$, $\phi_{A}$ in QCDF \cite{Cheng:2009cn} and     replaced by the long-distance charming penguins
    as $A_{cc}^{PP}$, $A_{cc}^{PV}$ and $A_{cc}^{VP}$ in
$B \to PP$, $B\to VP$ and $B\to PV$ decays, respectively  in SCET \cite{Wang:2008rk}.
Numerically it is not small. However, if one read this diagram carefully, one can find that it is not distinguishable in weak interaction from the diagram P in Fig.\ref{Penguin}(a).  The only difference between these two diagrams is the gluon exchange. Since all the QCD dynamics will be determined by $\chi^2$ fit from the experimental data, we will not introduce more parameters for this diagram in $B \to PP$ and $B\to VP$ decays.
The contribution of this diagram is already encoded in the parameter $r_{\chi}\chi^{P}$, $\mathrm{e}^{i\phi^{P}}$ in Eq.(\ref{eq:P}) for diagram Fig.\ref{Penguin}(a).
But for $B\to PV$ decays, we do need two
  parameters     $\chi^{P_A}$, $\mathrm{e}^{i\phi^{P_A}}$ for penguin annihilation diagram $P_A$ shown in Fig.\ref{Penguin}(d):
\begin{align}\label{eq:PA}
P_{A}^{PV}&=-\sqrt{2}G_{F}V_{tb}V_{tq^{'}}^{*}\chi^{P_{A}}\mathrm{e}^{i\phi^{P_{A}}}
f_{B}m_{V}(\frac{f_{P}f_{V}}{f_{\pi}^{2}})(\varepsilon^{*}_{V}\cdot p_{B}).
\end{align}

The contribution from $P_E$ diagram shown in Fig.\ref{Penguin}(c) is argued smaller than $P_A$ diagram, which
can be ignored reliably in decay modes not dominated by it such as
measured $B^{0}\to \pi^{+}\pi^{-}$, $B^{0}\to \pi^{0}\pi^{0}$,
$B^{0}\to K^{0}\bar{K^{0}}$ and $B^{0}\to \pi^{0}\rho^{0}$ decays. This $P_E$ contribution actually is the dominant contribution for the recent measurement of $B_s \to \pi^+ \pi^-$ decay \cite{Agashe:2014kda}
\begin{equation}
Br(B_s \to \pi^+ \pi^-) = (0.76\pm 0.19 ) \times 10^{-6} .\label{bspipi}
\end{equation}
We do not intend to use this single measurement to determine the contribution from  this diagram $P_E$. Thus we have to ignore it for later discussion.

The flavor-singlet QCD penguin diagram $P_C$  only contribute to the isospin singlet mesons $\eta$, $\eta'$, $\omega$ and $\phi$.  Anomaly related or not, there is also significant difference between these pseudo-scalar mesons and vector mesons.  We  distinguish  them as
$\chi^{P_C}$, $\mathrm{e}^{i\phi^{P_C}}$ for $B \to PP$ and
$B \to VP$ decays and $\chi^{P_C^{\prime}}$,
$\mathrm{e}^{i\phi^{P_C^{\prime}}}$ for $B \to PV$decays, respectively:
\begin{align}\label{eq:PC}
P_{C}^{PP}&=-i\frac{G_{F}}{\sqrt{2}}V_{tb}V_{tq^{'}}^{*}\chi^{P_C}\mathrm{e}^{i\phi^{P_C}}
 f_{p_{2}}(m_{B}^{2}-m_{p_{1}}^{2})F_{0}^{BP_{1}}(m_{p_{2}}^{2}) ,\nonumber\\
P_{C}^{PV}&=-\sqrt{2}G_{F}V_{tb}V_{tq^{'}}^{*} \chi^{P_C^{\prime}}\mathrm{e}^{i\phi^{P_C^{\prime}}}
f_{V}m_{V}F_{1}^{B-P}(m_{V}^{2})(\varepsilon^{*}_{V}\cdot p_{B}),\nonumber\\
P_{C}^{VP}&=-\sqrt{2}G_{F}V_{tb}V_{tq^{'}}^{*} \chi^{P_C}\mathrm{e}^{i\phi^{P_C}}
f_{P}m_{V}A_{0}^{B-V}(m_{P}^{2})(\varepsilon^{*}_{V}\cdot p_{B}),
\end{align}

The EW-penguin contribution is much smaller than QCD penguin diagram,
as the coupling coefficient $\alpha_{em}$ is one order smaller
than $\alpha_{s}$. We only keep its largest contribution diagram
shown in the second one of Fig.\ref{Penguin}, with gluon $g$
replaced by $Z$ or $\gamma$ with respect to QCD penguin diagram. Although the topology of $P_C$ diagram is quite similar to the $P_{EW}$ topology, their contributions are different. They both contribute to the isospin singlet meson emission decays. But $P_{EW}$ topology also contribute to the neutral isospin 1 meson emission decays. The topology of this diagram is very similar to the $T$ diagram. Factorization can be approved without ambiguity. Without introducing new parameters, we evaluate it
similar to $T$,
\begin{align}\label{eq:PEW}
P_{EW}^{PP}&=-i\frac{G_{F}}{\sqrt{2}}V_{tb}V_{tq^{'}}^{*}e_{q}\frac{3}{2}a_{9}(\mu)
f_{p_{2}}(m_{B}^{2}-m_{p_{1}}^{2})F_{0}^{BP_{1}}(m_{p_{2}}^{2}),\nonumber\\
P_{EW}^{PV}&=-\sqrt{2}G_{F}V_{tb}V_{tq^{'}}^{*}e_{q}\frac{3}{2}a_{9}(\mu)
f_{V}m_{V}F_{1}^{B-P}(m_{V}^{2})(\varepsilon^{*}_{V}\cdot p_{B}),\nonumber\\
P_{EW}^{VP}&=-\sqrt{2}G_{F}V_{tb}V_{tq^{'}}^{*}e_{q}\frac{3}{2}a_{9}(\mu)
f_{P}m_{V}A_{0}^{B-V}(m_{P}^{2})(\varepsilon^{*}_{V}\cdot p_{B}),
\end{align}
where $a_9(\mu)$ is the effective Wilson coefficient
  equal to $-0.009$ at scale $\mu=$2.1$\mathrm {GeV}$.

With all the decay amplitudes settled, the decay width for two-body charmless B decays is given by
\begin{align}
\Gamma(B\to M_{1}M_{2})=\frac{p}{8\pi m_B^2}\sum_{pol}|\mathcal{A}|^2 ,
\end{align}
where $M_{1}$, $M_{2}$ represent either two pseudoscalar $P_1$,$P_2$ or
one pseudoscalar $P$ and one vector $V$ in the final states. $p$ is
the 3 dimension momentum of either meson in the final state in
the center-of-mass frame. The summation over the polarization states
is for vector meson state.
The corresponding branching fraction is
\begin{align}
\mathcal{B}(B\to M_{1}M_{2})=\frac{\Gamma(B\to M_{1}M_{2})+
\Gamma(\bar{B}\to \bar{M_{1}} \bar{M_{2}})}{2}\times \tau_{B},
\end{align}
where $\tau_{B}$ is the $B$ meson lifetime.
The $CP$ violation charge asymmetry of exclusive $B^{-}$ and $B^{+}$ decay
is defined as
\begin{align}
\mathcal{A}_{cp}=\frac{\mathcal{B}(B^-\to \bar{M_{1}}\bar{M_{2}})-
\mathcal{B}(B^+\to M_{1}M_{2})}{\mathcal{B}(B^-\to \bar{M_{1}}\bar{M_{2}})+
\mathcal{B}(B^+\to M_{1}M_{2})}.\label{cp}
\end{align}
For the neutral $B_{(s)}$  mesons, there is a complication because of the  $B_{(s)}^{0}-\overline{B}_{(s)}^{0}$ mixing, if the decay product is a CP eigenstate. The     $CP$ asymmetry is  time dependent:
\begin{align}
\mathcal{A}_{cp}(t)=\mathcal{S}_{f}\mathrm {sin}(\Delta m_{B} t)-
\mathcal{C}_{f}\mathrm {cos}(\Delta m_{B} t),
\end{align}
where $\Delta m_{B}$ is the mass difference between the two mass eigenvalues
of B mesons. $\mathcal{A}_{cp} \equiv - \mathcal{C}_{f}$  is the direct $CP$ asymmetry and
$\mathcal{S}_{f}$ is the mixing induced
$CP$ asymmetry parameter, which are calculated as:
\begin{align}
\mathcal{C}_{f}&=\frac{1-|\lambda_f|^2}{1+|\lambda_f|^2},\nonumber\\
\mathcal{S}_{f}&=\frac{2Im(\lambda_f)}{1+|\lambda_f|^2} ,
\end{align}
where $\lambda_f=\frac{q}{p}\frac{\bar{A}_f}{A_f}$ and
$\frac{q}{p}=\frac{V^*_{tb}V_{td}}{V_{tb}V^*_{td}}$ or
$\big( \frac{V^*_{tb}V_{ts}}{V_{tb}V^*_{ts}}\big)$,
which is the mixing parameter for $B_{(s)}^{0}-\overline{B}_{(s)}^{0}$ mixing.
$A_f$ is the decay amplitude of $B^{0}\to f_{CP}$ and
$\bar{A}_f$ is the amplitude of the CP-conjugate process.

If the decay product is not a CP eigenstate, the $B_{(s)}^{0}-\overline{B}_{(s)}^{0}$ mixing will not result in a mixing induced CP asymmetry, but only a direct CP asymmetry like the $B^\pm$ decays (for example $B^0 \to \pi^-K^+$).
However, for the $\bar{B^0} \to \pi^\pm \rho^\mp$, $\bar{B^0} \to K_s \bar{K^{*0}}(K^{*0})$, $\bar{B_s}\to K^\pm K^{*\mp}$
and $\bar{B_s} \to K_s \bar{K^{*0}}(K^{*0})$ decay modes, the $B_{(s)}^{0}-\overline{B}_{(s)}^{0}$ mixing still plays an important role, even if the final states are not CP eigenstates. The reason is that both $B_{(s)}^{0}$ and $\overline{B}_{(s)}^{0}$ meson can decay to the same final state. The CP asymmetry is time dependent with four equations   \cite{Rui:2011dr}. There is a mismatch between theoretical and experimental variables. We adopt the convention of ref. \cite{Cheng:2014rfa}, for example, the mixing-induced $CP$ asymmetries $S_{cp}$
 for the $\bar{B^0} \to \pi^\pm \rho^\mp$ shown as,
\begin{align}
\mathcal{S}_{\bar{B^0} \to \pi^{+} \rho^{-}}&=\frac{2Im(\lambda_{\bar{B^0} \to \pi^{+} \rho^{-}})}{1+|\lambda_{\bar{B^0} \to \pi^{+} \rho^{-}}|^2} ,\nonumber\\
\mathcal{S}_{\bar{B^0} \to \pi^{-} \rho^{+}}&=\frac{2Im(\lambda_{\bar{B^0} \to \pi^{-} \rho^{+}})}{1+|\lambda_{\bar{B^0} \to \pi^{-} \rho^{+}}|^2} ,
\end{align}
where
\begin{align}
\lambda_{\bar{B^0} \to \pi^{+} \rho^{-}}&=\frac{q}{p} \frac{A(\bar{B^{0}} \to \pi^{+} \rho^{-})}{A(B^{0} \to \pi^{-} \rho^{+} )} ,\nonumber\\
\lambda_{\bar{B^0} \to \pi^{-} \rho^{+}}&=\frac{q}{p}\frac{A(\bar{B^0} \to \pi^{-} \rho^{+})}{A(B^0 \to \pi^{+} \rho^{-})}.
\end{align}
The definition of $S_{cp}$ for the $\bar{B^0} \to K_s \bar{K^{*0}}(K^{*0})$, $\bar{B_s}\to K^\pm K^{*\mp}$
and $\bar{B_s} \to K_s \bar{K^{*0}}(K^{*0})$ decays are similar with $\bar{B^0} \to \pi^\pm \rho^\mp$.

\section{Numerical Results and Discussions}\label{results}

\subsection{Input parameters}

The input parameters used in decay amplitudes mainly contain
the CKM matrix elements, decay constants and transition form factors.
We use the Wolfenstein parametrization for $V_{CKM}$ with the
    Wolfenstein parameters obtained from \cite{Agashe:2014kda}:
$$ \lambda=0.22537\pm 0.00061,~~~A=0.814^{+0.023}_{-0.024}$$
$$ \bar{\rho}=0.117\pm 0.021,~~~\bar{\eta}=0.353\pm 0.013.$$

Table \ref{tab:decay constants} represents the decay constants of
light meson ($P$, $V$). The measured $f_{\pi}$ and $f_{K}$ are given
in average by PDG \cite{Agashe:2014kda}. The value of $f_{B}$,$f_{B_{s}}$ and
the decay constants of vector mesons not measured directly in experiments
but can be got from several theoretical approaches, such as in
Covariant light front approach \cite{Cheng:2003sm}
light-cone sum rules \cite{Ball:2006eu,Straub:2015ica},
QCD sum rules \cite{Jamin:2001fw,Gelhausen:2013wia,
Penin:2001ux,Narison:2001pu,Lucha:2010ea,Narison:2012xy},
or lattice QCD \cite{Dowdall:2013tga,Carrasco:2013naa,Dimopoulos:2011gx,McNeile:2011ng,
Bazavov:2011aa,Na:2012kp,Bussone:2014bya,Bernardoni:2014fva}.
We show only central  values   in Table \ref{tab:decay constants}
and keep $5\%$ uncertainty, when estimate theoretical uncertainty of branching fractions and CP asymmetry parameters.

\begin{table}
\caption{The decay constants of light pseudo-scalar mesons  and vector mesons  (in unit of MeV).}\label{tab:decay constants}
\begin{center}
\begin{tabular}{cccccccccc}
\hline\hline
$f_{\pi}$ & $f_{K}$  & $f_{B}$ & $f_{B_{s}}$ & $f_{\rho}$ & $f_{K^{*}}$ & $f_{\omega}$ &$f_{\phi}$&
\\\hline
130 & 156 &190 & 225  & 213 & 220 &192 &225&
\\
\hline\hline
\end{tabular}
\end{center}
\end{table}

The transition form factors of $B$ meson decays were calculated
in various theoretical approaches, constitute quark model and light cone quark model
\cite{Melikhov:2000yu,Geng:2001de,Lu:2007sg,Albertus:2014bfa},
Covariant light front approach(LFQM) \cite{Cheng:2003sm,Cheng:2009ms,Chen:2009qk},
light-cone sum rules \cite{Ball:1998kk,Ball:2001fp,Ball:1998tj,Ball:2004ye,Ball:2004rg,Khodjamirian:2006st,Bharucha:2010im,Bharucha:2012wy,
Ball:2007hb,Charles:1998dr,Wu:2006rd,Duplancic:2008ix,Meissner:2013hya,Wang:2015vgv,Wu:2009kq,Khodjamirian:2011ub,Ivanov:2011aa,
Ahmady:2014sva,Fu:2014uea,Straub:2015ica},
 PQCD \cite{Li:2012nk,Wang:2012ab,Wang:2013ix,Fan:2013qz,
Fan:2013kqa,Kurimoto:2001zj,Lu:2002ny,Wei:2002iu,Huang:2004hw}
 and lattice QCD \cite{Horgan:2013hoa,Dalgic:2006dt,Aoki:2013ldr}.
The central values of the transition form factors
of $B$ meson decays at $q^{2}$=0 are shown in Table \ref{tab:formfactor}.
The error bar of
them are kept in $10\%$. This uncertainty of hadronic form factors is one of the major source of theoretical uncertainty in  our calculation  as shown in next section.
For the $q^{2}$ dependence of the transition form factors,  we use the dipole parametrization:
\begin{equation}\label{eq:ffdipole}
F_{i}(q^{2})={F_{i}(0)\over 1-\alpha_{1}{q^{2}\over M_{\rm pole}^{2}}+
\alpha_{2}{q^{4}\over M_{\rm pole}^{4}}},
\end{equation}
where $F_{i}$ denotes $F_{0}$, $F_{1}$, and $A_{0}$ in this article,
and $M_{\rm pole}$ is the mass of the corresponding pole state,
such as $B_{(s)}$ for $A_{0}$, and $B^{*}_{(s)}$ for $F_{0,1}$.
$\alpha_{1}$ and $\alpha_{2}$ are the dipole parameters
as shown in Table\ref{tab:formfactor}.
Since   the values      of  $q^{2}$=$\mathrm{m}^2_{P,V}$, where $\mathrm{m}_{P,V}$ is
the mass of emission meson in $B \to PP, PV$ decays, are small compared with the pole mass, this  $q^{2}$ dependence will not affect our numerical results significantly.

For the $\eta$ and $\eta^{\prime}$ meson in the final state of B decays,
their   decay constants and form factors are  defined through
  $\eta-\eta^{\prime}$ mixing,
\begin{equation}
\begin{pmatrix} \eta \\ \eta' \end{pmatrix}
=
\begin{pmatrix} \cos\phi & -\sin\phi \\ \sin\phi & \cos\phi \end{pmatrix}
\begin{pmatrix} \eta_{q} \\ \eta_{s} \end{pmatrix} ,
\end{equation}
where $\eta_{q}$ and $\eta_{s}$ are defined by
\begin{align}
\eta_{q}={1\over\sqrt2}(u\bar u+ d\bar d),~~~~~\eta_{s}=s\bar s.
\end{align}
The mixing angle is measured to be $\phi=(40.4\pm0.6)^{\circ}$ by
KLOE \cite{Ambrosino:2009sc}. The flavor decay constants of
$\eta_{q}$ and $\eta_{s}$ are $f_{q}=(1.07\pm0.02)f_{\pi}$ and
$f_{s}=(1.34\pm0.06)f_{\pi}$ respectively \cite{Feldmann:1998vh,Feldmann:1998sh}.
In a good approximation, we neglect the $\omega$ and $\phi$ mixing effect.

\begin{table}
\caption{The transition form factors of $B$ meson decays at $q^{2}$=0
and dipole model parameters }\label{tab:formfactor}
\centering
\begin{tabular}{|c||c|c|c|c|c||c|c|c|c|c|c|c|}
\hline

\hline
           &$~~~F_{0}^{B \to\pi}~~~$ &$~~~F_{0}^{B\to K}~~~$ &
           $~~~F_{0}^{B_{s}\to K}~~~$ & $~~~F_{0}^{B \to\eta_{q}}~~~$
           & $~~~F_{0}^{B_{s} \to\eta_{s}}~~~$ \\

\hline
$F(0)$     & 0.28 & 0.31  & 0.25  & 0.21  & 0.30   \\
$\alpha_1$ & 0.50 & 0.53  & 0.54  & 0.52  & 0.53    \\
$\alpha_2$ & -0.13 & -0.13 &-0.15  &   0    &   0     \\

\hline
           &$~~~F_{1}^{B \to\pi}~~~$ &$~~~F_{1}^{B\to K}~~~$ &
           $~~~F_{1}^{B_{s}\to K}~~~$ & $~~~F_{1}^{B \to\eta_{q}}~~~$
           & $~~~F_{1}^{B_{s} \to\eta_{s}}~~~$  \\

\hline
$F(0)$     & 0.28 & 0.31  & 0.25  & 0.21  & 0.30   \\
$\alpha_1$ & 0.52 & 0.54 & 0.57  & 1.43  & 1.48   \\
$\alpha_2$ & 0.45 & 0.50 & 0.50  &  0.41 & 0.46   \\
\hline

           &$~~~A_{0}^{B \to\rho}~~~$ &$~~~A_{0}^{B\to \omega}~~~$ &
           $~~~A_{0}^{B\to K^{*}}~~~$ & $~~~A_{0}^{B_{s} \to K^{*}}~~~$
           & $~~~A_{0}^{B_{s} \to\phi}~~~$ \\

\hline
$A(0)$     & 0.36 & 0.32  & 0.39  & 0.33  & 0.40   \\
$\alpha_1$ & 1.56 & 1.60  & 1.51  & 1.74  & 1.73    \\
$\alpha_2$ & 0.17 & 0.22 &0.14 &   0.47   &  0.41     \\
\hline
\hline
\end{tabular}
\end{table}

\subsection{The $\chi^{2}$ fit for  theoretical parameters  }

After the usage of factorization theorem, the number of theoretical parameters to be fitted from experimental data is reduced. The 6 parameters for tree diagrams are: color suppressed tree diagram amplitude $\chi^C$, $\chi^{C^{\prime}}$ and their phases $\phi^{C}$, $\phi^{C^{\prime}}$; W-exchange diagram amplitude $\chi^E$ and its phase $\phi^E$.
The 8 parameters for penguin diagrams are: Chiral enhanced penguin amplitude $\chi^P$ and its phase $\phi^P$; flavor singlet penguin amplitude $\chi^{P_C}$, $\chi^{P_C^{\prime}}$ and their phases $\phi^{P_C}$, $\phi^{P_C^{\prime}}$ for the pseudo-scalar and vector meson emission, respectively; the penguin annihilation amplitude $\chi^{P_A}$ and its phase $\phi^{P_A}$ for the vector meson emission only. Many of the charmless B decays channels have been experimentally measured \cite{Agashe:2014kda}. But some of them are measured with very poor precision. In our $\chi^2$ fit program, we will not use those data with less than $3\sigma$ significance. For the $B_s$ meson decays, very few modes are  measured, some of which are measured only by hadronic colliders such LHCb and CDF experiments. The precision of these $B_s$ decays measurements rely heavily on other $B$ decay channels measured by $B$ factories. Thus the systematic uncertainty of them is correlated. We will not use the $B_s$ decay data to avoid complications.
After these considerations, we have 37 branching Ratios and
11 CP violation observations of $B_{u,d} \to PP, PV$ decays from the current experimental data, where the branching ratios of $B^0 \to \pi^+ \rho^-$ and $B^0 \to \pi^- \rho^+ $ are derived from experimental data in ref.\cite{Cheng:2014rfa}.
With these 48 data, we use the  $\chi^{2}$ fit method by
Miniut program \cite{James:2004}
 to give the best-fitted parameters and
the corresponding $1\sigma$ uncertainty as:
\begin{align}\label{parameter}
\chi^{C}=0.48 \pm 0.06,~~~&\phi^{C}=-1.58 \pm 0.08,\nonumber \\
\chi^{C^{\prime}}=0.42 \pm 0.16,~~~&\phi^{C^{\prime}}=1.59\pm 0.17,\nonumber \\
\chi^{E}=0.057\pm0.005,~~~&\phi^{E}=2.71\pm 0.13,\nonumber\\
\chi^{P}=0.10\pm0.02,~~~&\phi^{P}=-0.61\pm 0.02.\nonumber \\
\chi^{P_C}=0.048 \pm 0.003,~~~&\phi^{P_C}=1.56 \pm 0.08,\nonumber \\
\chi^{P_C^{\prime}}=0.039\pm 0.003,~~~&\phi^{P_C^{\prime}}=0.68 \pm 0.08,\nonumber \\
\chi^{P_A}=0.0059\pm0.0008,~~~&\phi^{P_A}=1.51\pm 0.09,
\end{align}
with $\chi^{2}/\text{d.o.f}=45.2/34=1.3$.   This $\chi^{2}$ per degree of freedom is smaller than
the conventional flavor diagram approach
based on flavor $SU(3)$ symmetry\cite{Cheng:2014rfa}. In fact, they have  much more parameters   \cite{Cheng:2014rfa} than in our work.
 From eq.(\ref{parameter}), one can see that the color suppressed tree diagram amplitude $\chi^C$ and $\chi^{C^{\prime}}$ have similar size but  their phases $\phi^{C}$ and $\phi^{C^{\prime}}$ differ significantly.  Denoting  the pseudo-scalar and vector meson emission, respectively, their differences agree with the Glauber gluon effects \cite{Li:2014haa}. The similar differences are observed in the
flavor singlet penguin amplitude $\chi^{P_C}$, $\chi^{P_C^{\prime}}$ and their phases $\phi^{P_C}$, $\phi^{P_C^{\prime}}$ for the pseudo-scalar and vector meson emission, respectively.

To show the relative size of  every topological diagram  in each decay mode,
we take decay modes $B \to \pi \pi $ and $B \to \pi \rho $
to show the hierarchy of various tree and penguin topologies amplitude
($C$($P_C$) and $C^{\prime}$($P_C^{\prime}$) denote for the pseudo-scalar and vector meson emission respectively.),
 as follows:
\begin{align}
  T^{\pi \pi}:C^{\pi \pi}:E^{\pi \pi}:P^{\pi \pi}&=1:0.47:0.29:0.32 \label{hierarchy1} \\
  T^{\rho\pi }:C^{{\prime}^{\pi \rho}}:P^{\rho \pi}:P_{EW}^{\pi \rho}&=1:0.54:0.25:0.04\label{hierarchy11} \\
  T^{\pi\rho}:C^{\rho \pi }:P^{\rho \pi }:P_{EW}^{\rho \pi }&=1:0.36:0.19:0.03\label{hierarchy111}.
\end{align}
In these tree dominant decays, the relative importance of topological diagrams is easy to be reached:
\begin{equation}
T > C(C^\prime)> E \sim P >P_{EW}.
\end{equation}
This is in agreement with those QCD inspired approaches. For $B \to \pi K $
and $B \to \pi K^{*} $ decays, we have
  \begin{align}
  T^{\pi K}:C^{\pi K}:P^{\pi K}:P_{EW}^{\pi K}&=1:0.4:6.0:0.6\label{hierarchy2} \\
  T^{\pi K^{*}}:C^{K^{*}\pi }:P^{\pi K^{*}}:P_A^{\pi K^{*}}:P_{EW}^{K^{*} \pi }&=
  1:0.37:2.87:1.44:0.52\label{hierarchy22} .
\end{align}
In these penguin dominant decays, the relative importance of topological diagrams is also  reached  as:
\begin{equation}
P>P_A   >T >  P_{EW} >  C.
\end{equation}
It is interesting that the electroweak penguin contribution $P_{EW}$ is even more larger than
the color suppressed tree $C$. It is indeed not negligible. For $B \to \rho K $   decays, we have
\begin{equation}
  T^{\rho K }:C^{{\prime}^{K \rho}}:P^{\rho K }:P_{EW}^{K \rho} =1:0.49:2.82:0.79  \label{hierarchy222}.
  \end{equation}
In this channel, we have very similar contributions from each topology:
\begin{equation}
P  >T >  P_{EW}> C^\prime.
\end{equation}
Again, the electroweak penguin contribution $P_{EW}$ is important.

As the $P_C$ and $P_C^{\prime}$ only contribute to
modes including flavor singlets   meson ($\eta,\eta^{\prime},\omega,\phi$),
the hierarchy including $P_C$ and $P_C^{\prime}$ are represented as:
\begin{align}
 T^{\pi \eta}:C^{\pi \eta}:P^{\pi \eta}:P_C^{\pi \eta}:P_{EW}^{\pi \eta}&=1:0.50:0.57:0.06:0.03\label{hierarchy3} \\
 T^{\eta K}:C^{\eta K}:P^{\eta K}:P_C^{\eta K}:P_{EW}^{\eta K}&=1:0.45:3.39:1.10:0.52\label{hierarchy33}\\
 T^{\pi \omega}:C^{{\prime}^{\pi \omega}}:P^{\pi \omega}:P_C^{{\prime}^{\pi \omega}}:
 P_A^{\pi \omega}:P_{EW}^{\pi \omega}&=1:0.54:0.21:0.26:0.10:0.02\label{hierarchy333}
\end{align}
The flavor singlet penguin contribution $P_C$ is important, as it is even larger than the color favored tree contribution $T$ in $\eta K$ channel and it is at the similar  size as penguin emission contribution $P$ in $\pi \omega$ channel.
The importance of this type of penguin contribution is recently emphasized \cite{he}.

\subsection{Branching Ratios for the charmless B decays}

\begin{table}[htbp]
\caption{Branching fractions $(\times10^{-6})$ of various $\bar B \to PP$ decay modes.
  We also show the   experimental data   \cite{Agashe:2014kda} and  results from conventional flavor diagram approach \cite{Cheng:2014rfa}   for comparison. }\label{tab:BrBudPP}
\begin{center}
\footnotesize\begin{tabular}{cccccccc}
\hline\hline
  & Mode & Amplitudes & ~~~~$\mathcal{\rm Exp}$~~~~ &
~~~~This work~~~~&Flavor diagram   \\
\hline
&$ \pi^{-}  \pi^{0}$     & $T,C,P_{EW}$&               $\star 5.5\pm0.4$                              &$5.08\pm0.39\pm1.02\pm0.02$&$5.40\pm0.79$\\
&$ \pi^{-} \eta$         & $T,C,P,P_C,P_{EW} $&        $\star 4.02\pm0.27$                            &$4.13\pm0.25\pm0.64\pm0.01$& $3.88\pm0.39$\\
&$ \pi^{-} \eta^{'}$ &       $ T,C,P,P_C,P_{EW}  $& $\star 2.7\pm0.9$       &$3.37\pm0.21\pm0.49\pm0.01$&$5.59\pm0.54$\\
&$ \pi^{+} \pi^{-}$      & $ T,E,(P_E),P$&             $\star 5.12\pm0.19$                            &$5.15\pm0.36\pm1.31\pm0.14$& $5.17\pm1.03$\\
&$  \pi^{0}  \pi^{0}$    & $ C,E,P,(P_E),P_{EW}$&      $\star 1.91\pm0.22$                            &$1.94\pm0.30\pm0.28\pm0.05$& $1.88\pm0.42$\\
&$  \pi^{0} \eta$        & $C,E,P_C,(P_E),P_{EW} $&    $<1.5$                                         &$0.86\pm0.08\pm0.08\pm0.04$& $0.56\pm0.03$\\
&$  \pi^{0} \eta^{'}$    & $C,E,P_C,(P_E),P_{EW} $&    $1.2\pm0.6$                                    &$0.87\pm0.08\pm0.10\pm0.03$& $1.21\pm0.16$\\
&$ \eta \eta$            & $ C,E,P_C,(P_E),P_{EW}$&    $<1.0$                                         &$0.44\pm0.09\pm0.08\pm0.005$& $0.77\pm0.12$\\
&$ \eta \eta^{'}$        & $ C,E,P_C,(P_E),P_{EW}$&    $<1.2$                                         &$0.77\pm0.13\pm0.14\pm0.008$& $1.99\pm0.26$\\
&$ \eta^{'} \eta^{'}$    & $C,E,P_C,(P_E),P_{EW} $&    $<1.7$                                         &$0.38\pm0.05\pm0.07\pm0.003$&$1.60\pm0.20$\\
&$ K^{-} K^{0}$           & $ P $&                     $\star 1.31\pm0.17$                            &$1.32\pm0.04\pm0.26\pm0.01$&$1.03\pm0.02$\\
&$ K^{0} \bar{K^{0}}$     & $P $&                      $\star 1.21\pm0.16$                            &$1.23\pm0.03\pm0.25\pm0.01$&$0.89\pm0.11$\\
\hline
&$ \pi^{-} \bar{K^{0}}$   & $ P $&      $\star 23.7\pm0.8$                             &$23.2\pm0.6\pm4.6\pm0.2$&$23.53\pm0.42 $ \\
&$  \pi^{0} K^{-}$        & $T,C,P,P_{EW} $&           $\star 12.9\pm0.5$                             &$12.8\pm0.32\pm2.35\pm0.10$&$ 12.71\pm1.05$\\
&$ \eta K^{-}$ &   $ T,C,P,P_C,P_{EW}$&    $\star 2.4\pm0.4$                 &$2.0\pm0.13\pm1.19\pm0.03$&$1.93\pm0.31 $ \\
&$ \eta^{'} K^{-}$        & $T,C,P,P_C,P_{EW}$&     $\star 70.6\pm2.5$                                &$70.1\pm4.7\pm11.3\pm0.22$&$70.92\pm8.54 $ \\
&$ \pi^{+} K^{-}$         & $ T,P $&               $\star 19.6\pm0.5$                                 &$19.8\pm0.54\pm4.0\pm0.2$&$20.2\pm0.39 $\\
&$  \pi^{0} \bar{K^{0}}$  & $C,P,P_{EW}$&          $\star 9.9\pm0.5$                                  &$8.96\pm0.26\pm1.96\pm0.09$&$9.73\pm0.82 $\\
&$ \eta \bar{K^{0}}$      & $C,P,P_C,P_{EW} $&      $\star 1.23\pm0.27$                               &$1.35\pm0.10\pm1.02\pm0.03$&$1.49\pm0.27 $\\
&$ \eta^{'} \bar{K^{0}}$  & $C,P,P_C,P_{EW} $&      $\star 66\pm4$                                    &$66.4\pm4.5\pm10.6\pm0.21$&$66.51\pm7.97 $ \\
\hline\hline
\end{tabular}
\end{center}
\end{table}

After $\chi^2$ fitting the parameters in Eq.(\ref{parameter}), we get
the numerical results of branching fractions for
$\bar B\to PP$ decays   shown in Table~\ref{tab:BrBudPP} and
$\bar B\to PV$ decays in Table~\ref{tab:BrBudPV}.
Each branching fractions tables are divided into two parts:
$\Delta S=0$ transitions and $\Delta S=1$ transitions.
We also show the contributing topological amplitude symbols for each channel in these tables.
For the theoretical uncertainties in the tables (apply also to the following tables), the first one is
  the  statistical uncertainty
from the $\chi^2$ fitting by experimental data. The second one arise from
the transition form factors which are set to be $10\%$ uncertainties,
and the third from decay constants. We can find that the dominant
uncertainty for most channels is from form factors, which need to be approved
by theories and semi-leptonic B decay measurements. The experimental data are also shown in these tables to
compare with theoretical predictions. Not all of the measurements are in a good accuracy. In our $\chi^2$ fit  program, we use those data only with more than 3$\sigma$ signal significance that marked as a $*$ in these tables.
The rest can be considered as theoretical predictions,   waiting for LHCb and other experiments
to test.

From Table \ref{tab:BrBudPP} and \ref{tab:BrBudPV},  one can easily find that
$\mathrm{B}$($B^{-}\to \pi^{-}\pi^{0}$) is twice smaller  than
$\mathrm{B}$($B^{-}\to \pi^{0}\rho^{-}$).  These two modes receive similar contributions from the color favored tree diagram
denoted by $T$, while all other contributions are suppressed. If not considering  SU(3) breaking effects, one need two parameters to fit these two diagrams in ref.\cite{Cheng:2014rfa}.  In our FAT approach, this can be easily explained by the fact that $f_{\rho}>f_{\pi}$, therefore we do not need any free parameter to be fitted from experimental data. Due to the difference between vector or pseudo-scalar emission in color suppressed tree diagram
$\chi^{C^{\prime}}$ and  $\chi^{C}$, especially the very larger strong phase difference, the $\mathrm{B}$($B^{-}\to \rho^{-}\pi^{0}$)
is a little different from $\mathrm{B}$($B^{-}\to \rho^{0}\pi^{-}$).  Interestingly,
for the penguin dominated B decays it is the inverse situation.  The branching fractions of the penguin diagram $P$ dominated decay modes
$B^{-}\to \pi^{-} \bar{K^{0}}$, $B^{-}\to \pi^{0} K^{-}$ and
$\bar B^{0}\to \pi^{0} \bar{K^{0}}$ are larger than their
corresponding ones of $B\to PV$ decays.  This can be understood from eq.(\ref{eq:P}) that in addition to the factorizable amplitude of QCD penguin  emission topology, there is a large chiral enhanced penguin contribution
 in $B \to PP$ modes; while no such contribution in $B\to PV$ modes and negative contribution in $B\to VP$ modes.

\begin{table}[htbp]
\caption{Branching fractions $(\times10^{-6})$ of various $\bar B \to PV$ decay modes.
  We also show   the experimental data     \cite{Agashe:2014kda} and
  results from conventional flavor diagram approach \cite{Cheng:2014rfa}   for comparison.}
\label{tab:BrBudPV}
\begin{center}
\scriptsize\begin{tabular}{ccccccc}
\hline\hline
  & Mode & Amplitudes & ~~~~$\mathcal{\rm Exp}$~~~~ &
~~~~This work~~~~&Flavor diagram   \\
\hline
&$ \pi^{-} \rho^{0}$      & $T,C^{\prime},P,P_A,P_{EW}$               &$\star 8.3\pm1.2$       &$8.6\pm1.81\pm1.38\pm0.03$&$7.59\pm1.41$\\
&$ \pi^{-} \omega$        & $T,C^{\prime},P,P_C^{\prime},P_A,P_{EW} $   &$\star 6.9\pm0.5$    &$6.78\pm1.46\pm1.09\pm0.02$&$7.03\pm1.42$\\
 &$ \pi^{-} \phi$          & $P_C^{\prime},P_{EW} $                  &$<0.15$     &$0.28\pm0.004\pm0.055\pm0.003$&$0.04\pm0.02$\\
&$  \pi^{0} \rho^{-}$     & $T,C,P,P_A,P_{EW}$                         &$\star 10.9\pm1.4$           &$12.9\pm0.73\pm2.30\pm0.12$&$12.15\pm2.52$\\
&$ \eta \rho^{-}$         & $T,C,P,P_C,P_A,P_{EW}$                      &$7.0\pm2.9$                   &$8.16\pm0.48\pm1.43\pm0.07$&$5.26\pm1.19$\\
&$ \eta^{'} \rho^{-}$     & $T,C,P,P_C,P_A,P_{EW}$                      &$\star 9.7\pm2.2$            &$6.0\pm0.34\pm0.97\pm0.05$&$5.66\pm1.25$\\
&$ \pi^{+} \rho^{-}$      & $T,E,P,(P_E),P_A $                          &$\star 14.6\pm1.6 $            &$12.4\pm0.64\pm3.20\pm0.38$&$15.20\pm1.52$\\
&$ \pi^{-} \rho^{+}$      & $ T,E,P,(P_E)$                               &$\star 8.4\pm1.1$               &$6.04\pm0.47\pm1.70\pm0.25$&$8.22\pm1.06$\\
&$  \pi^{0} \rho^{0}$     & $C,C^{\prime},E,P,P_A,(P_E),P_{EW}$              &$\star 2\pm0.5$           &$1.32\pm0.47\pm0.09\pm0.14$&$2.24\pm0.93$\\
&$  \pi^{0} \omega$       & $C,C^{\prime},E,P,P_A,(P_E),P_{EW} $             &$<0.5$                      &$2.31\pm0.88\pm0.24\pm0.07$&$1.02\pm0.66$\\
&$  \pi^{0} \phi$         & $ P_C^{\prime},P_{EW}$                  &$<0.15$                           &$0.13\pm0.002\pm0.025\pm0.001$&$0.02\pm0.01$\\
&$ \eta \rho^{0}$         & $C,C^{\prime},E,P,P_C,P_C^{\prime},P_A,(P_E),P_{EW} $        &$ <1.5 $      &$4.41\pm1.15\pm0.39\pm0.17$ & $0.54\pm0.32$\\
&$ \eta \omega$           & $C,C^{\prime},E,P,P_C,P_C^{\prime},P_A,(P_E),P_{EW}$         &$0.94^{+0.40}_{-0.31}$&$0.89\pm0.30\pm0.08\pm0.09$&$1.12\pm0.44$\\
&$ \eta \phi$             & $P_C^{\prime},P_{EW} $                            &$<0.5$                    &$0.077\pm0.001\pm0.015\pm0.0008$&$0.01\pm0.01$\\
&$ \eta^{'} \rho^{0}$     & $C,C^{\prime},E,P,P_C,P_C^{\prime},(P_E),P_{EW} $        &$<1.3$             &$3.19\pm0.77\pm0.29\pm0.12$&$0.63\pm0.33$\\
&$ \eta^{'} \omega$       & $C,C^{\prime},E,P,P_C,P_C^{\prime},(P_E),P_{EW}$         &$1.0^{+0.5}_{-0.4}$  &$0.95\pm0.21\pm0.05\pm0.06$&$1.24\pm0.47$\\
&$ \eta^{'} \phi$         & $ P_C^{\prime},P_{EW}$                            &$<0.5$                       &$0.05\pm0.0008\pm0.01\pm0.0005$&$0.01\pm0.01$\\
&$ K^{-} K^{*0}$                & $ P,P_A$&                                $<1.1$                        &$0.59\pm0.06\pm0.10\pm0.01$&$0.46\pm0.03$\\
&$ K^{0} K^{*-}$                & $ P $&                                  $     $                          &$0.44\pm0.03\pm0.09\pm0.004$&$0.31\pm0.03$\\
&$ K^{0} \bar{K^{*0}}$          & $P $&                                   $     $                         &$0.41\pm0.02\pm0.08\pm0.004$&$0.29\pm0.03$\\
&$ \bar{K^{0}} K^{*0}$          & $P,P_A$&                                 $     $                           &$0.55\pm0.05\pm0.09\pm0.01$&$0.43\pm0.02$\\
\hline
  &$ \pi^{-} \bar{K^{*0}}$        & $ P,P_A $&                $\star 10.1\pm0.9$             &$10.0\pm0.95\pm1.78\pm0.15$&$10.47\pm0.60$\\
&$  \pi^{0} K^{*-}$             & $T,C,P,P_A,P_{EW} $&                     $\star 8.2\pm1.9$                  &$6.23\pm0.51\pm0.98\pm0.07$&$9.79\pm2.95$\\
 &$ \eta K^{*-}$         & $T,C,P,P_C,P_A,P_{EW}$&     $\star 19.3\pm1.6$      &$17.3\pm0.8\pm2.4\pm0.3$&$16.57\pm2.58$\\
&$ \eta^{'} K^{*-}$             & $T,C,P,P_C,P_A,P_{EW}  $&                 $4.8^{+1.8}_{-1.6}$            &$3.31\pm0.44\pm0.38\pm0.13$&$3.43\pm1.43$\\
&$ K^{-} \rho^{0}$              & $T,C^{\prime},P,P_{EW}$&                $\star 3.7\pm0.5$             &$3.97\pm0.25\pm0.80\pm0.04$&$3.97\pm0.90$\\
&$ K^{-} \omega$                & $T,C^{\prime},P,P_C^{\prime},P_{EW}$&    $\star 6.5\pm0.4$               &$6.52\pm0.73\pm1.13\pm0.06$&$6.43\pm1.49$\\
&$ K^{-} \phi$                  & $P,P_C^{\prime},P_A,P_{EW} $&             $\star 8.8\pm0.7$             &$8.38\pm1.21\pm0.69\pm0.50$&$8.34\pm1.31$\\
&$ \bar{K^{0}} \rho^{-}$        & $ P $&                                  $\star 8\pm1.5$               &$7.74\pm0.47\pm1.55\pm0.07$&$7.09\pm0.77$\\
&$ \pi^{+} K^{*-}$              & $ T,P,P_A$&                              $\star 8.4\pm0.8$              &$8.40\pm0.77\pm1.46\pm0.14$&$8.35\pm0.50$\\
&$  \pi^{0} \bar{K^{*0}}$       & $ C,P,P_A,P_{EW}$&                       $\star 3.3\pm0.6$           &$3.35\pm0.36\pm0.65\pm0.08$&$3.89\pm1.98$\\
&$ \eta \bar{K^{*0}}$           & $C,P,P_C,P_A,P_{EW} $&                    $\star 15.9\pm1$            &$16.6\pm0.7\pm2.3\pm0.3$&$16.34\pm2.48$\\
&$ \eta^{'} \bar{K^{*0}}$       & $C,P,P_C,P_C^{\prime},P_A,P_{EW} $&        $\star 2.8\pm0.6$          &$3.0\pm0.5\pm0.3\pm0.1$&$3.14\pm1.24$\\
&$ K^{-} \rho^{+}$              & $T,P $&                                 $\star 7\pm0.9$                &$8.27\pm0.44\pm1.65\pm0.07$&$8.28\pm0.80$\\
&$ \bar{K^{0}} \rho^{0}$        & $C^{\prime},P,P_{EW} $&                 $\star 4.7\pm0.4$              &$4.59\pm0.34\pm0.79\pm0.04$&$4.97\pm1.14$\\
&$ \bar{K^{0}} \omega$          & $C^{\prime},P,P_C^{\prime},P_{EW} $&     $\star 4.8\pm0.6$            &$4.80\pm0.61\pm0.95\pm0.05$&$4.82\pm1.26$\\
&$ \bar{K^{0}} \phi$            & $P,P_C^{\prime},P_A,P_{EW}$&              $\star 7.3\pm0.7$             &$7.77\pm1.12\pm0.64\pm0.46$&$7.72\pm1.21$\\
\hline\hline
\end{tabular}
\end{center}
\end{table}

Similar to the conventional topological diagram approach \cite{Cheng:2014rfa},
   the long-standing puzzle of  large
 $B^0\to \pi^{0}\pi^{0}$  branching fraction   can be resolved well attributed to
the appropriate magnitude and phase of $C$ in FAT.     Naive estimation indicates that
$\vert$C$\vert/\vert$T$\vert$ is about 1/3 due to color suppressed factor.
The $\chi^{C}$ are enhanced by large nonperturbative contribution such as
final states interaction and re-scattering effects. Although some power
corrections to them were parameterized in QCDF, PQCD and SCET as
mentioned before, where the $\pi\pi$ puzzle was accommodated to
some extent, it is not resolved completely in those factorization approaches.
With a larger $B^0\to \pi^{0}\pi^{0}$  branching fraction, the $B^0\to \pi^0 \rho^0$ and $B^0\to \rho^0 \rho^0$
branching ratio will go easily much larger than the experimental data.
Actually,  only the Glauber phase
factor \cite{Li:2014haa}, associated with the Goldstone boson  $\pi$ can resolve
the $B\to\pi\pi$, $B\to \pi\rho$ and $B \to \rho\rho$ puzzles consistently.
We   predict these branching ratios correctly in Table \ref{tab:BrBudPP}
with not too large $\chi^{C}$.
$|T^{\pi \pi}|:|C^{\pi \pi}|=1:0.47$ shown in Eq.(\ref{hierarchy1})
is not as large as \cite{Cheng:2014rfa}, where the ratio even reached 0.97
in Scheme C.

The $B^{-}\to K^{-}K^{0}$, $B^{0}\to K^{0}\bar{K^{0}}$ decays are purely penguin decays.
 From Table \ref{tab:BrBudPP} one can see that their branching fractions  given in our FAT approach are in much better  agreement with experimental data than the previous conventional flavor diagram approach \cite{Cheng:2014rfa}. The   penguin amplitude is mostly determined by the more precise measurements of $B^0 \to \pi K$ decays.  There is only SU(3) breaking effect between $KK$ final states and $\pi K$ final states. Our results for $KK$ final sates are larger, because we considered SU(3) breaking effect and
    the previous conventional flavor diagram approach not. For the $B\to PV $ decays,
where a vector meson is emitted from the weak interaction point, such as
$B^{-}\to K^{-}K^{*0}$, $\bar B^{0}\to K^{*0}\bar{K^{0}}$ decay modes, there is an extra  penguin annihilation diagram $P_A$, in addition to the penguin emission
 diagram $P$.
We find that the theoretical prediction for $\mathrm{B}(B^{-}\to K^{-}K^{*0})$
 is a little larger than
$\mathrm{B}(B^{-}\to K^{*-}K^{0})$, and $\mathrm{B}(\bar B^{0}\to \bar K^{0}K^{*0})$
 is a little larger than
$\mathrm{B}(\bar B^{0}\to \bar K^{*0}K^{0})$. All these results are larger than the previous conventional flavor diagram approach \cite{Cheng:2014rfa}, but in agreement with the prediction from SCET \cite{Wang:2008rk}.

We did not show the decay   $ \bar B^{0} \to K^{+} K^{-}$ in our table. This decay is     measured
with   $\mathrm{B}(\bar B^{0} \to K^{+} K^{-}) =0.13\pm0.05$  that is less than
 $3\sigma$ significance, therefore, we did not include this measurement into our $\chi^2$ fit program. Theoretically, this decay is dominated by the exchange diagram $E$  and penguin exchange diagram $P_E$.
Since not enough experimental data to fit the    $P_E$ contribution, our result for this channel is only from    the W-exchange diagram $\chi^{E} $
fitted from $B^{0} \to \pi^{0}\pi^{0}(\rho^{0})$, $\pi^{+}\pi^{-}$ decay modes.    With only one contribution,  our result $ \mathrm{B}(B^{0} \to K^{+} K^{-}) =1.30\pm0.25\pm0.00\pm0.13$ is one order magnitude higher than the central value of experimental data.
 This should be resolved with more precise experimental data to fit the $P_E$ contribution in the future.

The $B \to PP $ decays with flavour singlet mesons $\eta^{(\prime)}$
in the final states are more complicated than other  decay channels.   There are complicated $\eta-\eta^{\prime}$
mixing effect and most of them   include almost all kinds of topologies
except for $E$ diagram. As shown in Eq.(\ref{hierarchy3}), $\vert P_C\vert/\vert P\vert$
is close to $\vert C\vert/\vert T\vert$ in $\eta K$ decays. The flavor-singlet QCD penguin  diagram $P_C$
in FAT approach and also in the conventional topological diagram approach \cite{Cheng:2014rfa} play the same role as  the
long-distance charming penguin $A_{ccg}^{PP},A_{ccg}^{VP}$ in SCET \cite{Wang:2008rk}.
It has an important effect on the large branching fraction of $B \to K \eta^{\prime}$
and other observations of this type of penguin dominant decays.
In    the conventional topological diagram approach,    the $\eta-\eta^{\prime}$
mixing angle $\phi$ is a free parameter to be fitted from hadronic B decay data as  $\phi=46^{0}$ for $B\to PP$ and
 $\phi=43^{0}$ for $B\to PV$ decays \cite{Cheng:2014rfa}.
 However, the fitting is not so successful as expected with the  branching faction  of $B^{-}\to \pi^{-} \eta^{\prime}$ two times
larger   than the experimental value. These decays are recently reanalyzed with better results for $B\to PP$ decays in ref.\cite{he}. It is noted that we fix the mixing angles from other experiments for the  $\eta-\eta^{\prime}$, resulting in better results for these decays.

\begin{table}[htbp]
\caption{Branching fractions $(\times10^{-6})$ of various $\bar B_{s}\to PP$ and $\bar B_{s}\to PV$ decays.
We also show the experimental data   \cite{Agashe:2014kda} and  results from conventional flavor diagram approach \cite{Cheng:2014rfa}  for comparison. }
\label{tab:BrBsPPPV}
\begin{center}
\scriptsize\begin{tabular}{ccccccc}
\hline\hline
  & Mode & Amplitudes & ~~~~$\mathcal{\rm Exp}$~~~~ &
~~~~This work~~~~&Flavor diagram   \\
\hline
&$ \pi^{-} K^{+}$           & $ T,P$&              $5.5\pm0.6$                              &$6.98\pm0.02\pm1.40\pm0.02$          &$5.86\pm0.78$\\
&$  \pi^{0} \eta$           & $ C,E,P_C,(P_E),P_{EW}$&    $<1000$                           &$0.10\pm0.013\pm0.013\pm0.003$         &$0.12\pm0.07$\\
&$  \pi^{0} \eta^{'}$       & $ C,E,P_C,(P_E),P_{EW} $&   $     $                           &$0.11\pm0.01\pm0.02\pm0.002$             &$0.12\pm0.06$\\
&$  \pi^{0} K^{0}$          & $C,P,P_{EW} $ &   $     $                                      &$0.97\pm0.16\pm0.2\pm0.003$              &$2.25\pm0.33$\\
&$ \eta \eta$               & $C,E,P,P_C,(P_E),P_{EW} $&  $<1500$                            &$11.4\pm0.42\pm2.25\pm0.04$          &$8.24\pm1.53$\\
&$ \eta \eta^{'}$           & $C,E,P,P_C,(P_E),P_{EW} $&  $     $                         &$40.4\pm2.06\pm8.14\pm0.13$            &$33.47\pm3.64$\\
&$ \eta K^{0}$              & $C,P,P_C,P_{EW} $&    $     $                                 &$0.55\pm0.11\pm0.08\pm0.002$           &$0.97\pm0.16$\\
&$ \eta^{'} \eta^{'}$       & $C,E,P,P_C,(P_E),P_{EW}$&   $     $                         &$42.1\pm3.48\pm8.38\pm0.13$            &$41.48\pm6.25$\\
&$ \eta^{'} K^{0}$          & $C,P,P_C,P_{EW}$&     $     $                              &$2.15\pm0.15\pm0.30\pm0.01$           &$3.94\pm0.39$\\
&$ K^{+} K^{-}$             & $T,E,P,(P_E) $&            $24.9\pm1.7$                    &$16.7\pm0.46\pm3.27\pm0.16$          &$17.90\pm2.98$\\
&$ K^{0} \bar{K^{0}}$       & $P $&                $<66  $                              &$17.5\pm0.47\pm3.50\pm0.16$          &$17.48\pm2.36$\\
\hline
&$ \pi^{-} K^{*+}$          & $T,P $&                  $ $                                         &$11.1\pm0.02\pm2.21\pm0.03$            &$7.92\pm1.02$\\
&$  \pi^{0} \phi$           & $C,P_{EW} $   &          $ $                                        &$0.26\pm0.02\pm0.05\pm0.001$           &$1.94\pm1.14$\\
&$  \pi^{0} K^{*0}$         & $C,P,P_{EW} $   &        $ $                                       &$1.22\pm0.25\pm0.24\pm0$          &$3.07\pm1.20$\\
&$ \eta \rho^{0}$           & $C^{\prime},E,P_C^{\prime},(P_E),P_{EW}$   &          $ $           &$0.13\pm0.02\pm0.02\pm0.003$           &$0.34\pm0.21$\\
&$ \eta \omega$             & $C^{\prime},E,P_C^{\prime},(P_E),P_{EW}  $   &        $ $            &$3.25\pm0.10\pm0.63\pm0.03$           &$0.15\pm0.16$\\
&$ \eta \phi$               & $C,P,P_C,P_C^{\prime},P_{EW},P_A$   &             $ $               &$0.80\pm0.22\pm0.53\pm0.14$            &$0.39\pm0.39$\\
&$ \eta K^{*0}$             & $C,P,P_C,P_{EW},P_A $   &                        $ $                 &$0.99\pm0.18\pm0.16\pm0.01$            &$1.44\pm0.54$\\
&$ \eta^{'} \rho^{0}$       & $ C^{\prime},E,P_C^{\prime},(P_E),P_{EW}$   &         $ $               &$0.37\pm0.07\pm0.05\pm0.01$           &$0.31\pm0.19$\\
&$ \eta^{'} \omega$         & $C^{\prime},E,P_C^{\prime},(P_E),P_{EW} $   &         $ $           &$3.97\pm0.15\pm0.79\pm0.04$           &$0.14\pm0.14$\\
&$ \eta^{'} \phi$           & $C,P,P_C,P_C^{\prime},P_{EW},P_A$   &             $ $                &$13.0\pm1.05\pm0.98\pm0.67$           &$5.48\pm1.84$\\
&$ \eta^{'} K^{*0}$         & $C,P,P_C,P_{EW},P_A  $   &                       $ $             &$1.64\pm0.15\pm0.22\pm0.03$          &$1.65\pm0.60$\\
&$ K^{+} \rho^{-}$          & $T,P,P_A $   &                                  $ $              &$17.5\pm0\pm3.5\pm0.2$           &$14.63\pm1.46$\\
&$ K^{+} K^{*-}$            & $ T,E,P,P_A,(P_E)$   &                                $ $             &$8.85\pm1.06\pm1.04\pm0.37$      &$8.03\pm0.48$\\
&$ K^{-} K^{*+}$            & $ T,E,P,(P_E)$   &                                   $ $               &$6.39\pm0.38\pm1.35\pm0.07$           &$7.98\pm0.77$\\
&$ K^{0} \rho^{0}$          & $ C^{\prime},P,P_C^{\prime},P_A,P_{EW}$   &      $ $                  &$1.61\pm1.10\pm0.31\pm0.02$           &$0.56\pm0.24$\\
&$ K^{0} \omega$            & $C^{\prime},P,P_C^{\prime},P_A,P_{EW} $   &      $ $                    &$1.43\pm0.88\pm0.25\pm0.02$          &$0.58\pm0.25$\\
&$ K^{0} \phi$              & $ P,P_C^{\prime},P_{EW}$   &                    $ $                  &$0.35\pm0.04\pm0.06\pm0.003$           &$0.41\pm0.07$\\
&$ K^{0} \bar{K^{*0}}$      & $P,P_A $   &                                    $ $                   &$9.28\pm1.14\pm1.21\pm0.34$           &$9.33\pm0.54$\\
&$ \bar{K^{0}} K^{*0}$      & $P $   &                                       $ $                     &$6.31\pm0.38\pm1.26\pm0.06$           &$6.32\pm0.68$\\

\hline\hline
\end{tabular}
\end{center}
\end{table}

For the sub-leading contribution electroweak penguin diagram $P_{EW}$,  four free parameters (two magnitudes and two phases) are introduced to be fitted from experiments \cite{Cheng:2014rfa} with non-negligible
 strong phase for  $B\to PP$ decays and
even considerable magnitude   for $B\to VP$ decays.  As stated in the last section, we did not  include any free parameters for this kind of diagrams but use
factorization formulas  to make predictions.   For the $B \to \pi(\rho) K(K^{*})$ decays, their branching fractions
are in good agreement with data   by the non-negligible factorization $P_{EW}$ diagram contribution.
For example, the central value of
$\mathrm{B}$($B^{-}\to \pi^{0} K^{-}$) is equal to
data precisely attributed to the non-negligible correction effect
from $P_{EW}$ diagram.

Most of the
$B_s \to PP$, $PV$ decays are not well measured in the experiments. Therefore, we do not include any of the $B_s$ data in our $\chi^2$ fit program. Their branching ratios are all as predictions in our FAT approach shown in Table \ref{tab:BrBsPPPV}. The accuracy of these predictions rely on the assumption that the mechanism for $B$ and $B_s$ decays are the same. If there are enough data for $B_s$ decays, one need do the $\chi^2$ fit again. In this table, we do not include the channel $B_{s}\to \pi^{+}\pi^{-}$.
Our result (with only W-exchange contribution) for this channel $ \mathrm{B} ( \bar B_{s}\to \pi^{+}\pi^{-}) =0.051\pm0.001\pm0\pm0.005$  is much
smaller than the experimental data measured   by LHCb  and CDF shown in Eq.(\ref{bspipi}). As stated in the last section, this decay is dominated by  the     penguin exchange diagram $P_E$ \cite{bspipi}, which can only be fitted from
this mode $B_{s}\to \pi^{+}\pi^{-}$. One measurement to determine one parameter is not a perfect way of $\chi^2$ fitting. Therefore we look forward
to more data to determine this contribution in other modes and to test our FAT in the future.
Similarly,     without this contribution, we are unable to predict a number of decay channels, dominated by this contribution: $B^0 \to K^+K^-$, $B^0 \to K^{*+}K^-$, $B^0 \to K^+K^{*-}$, $B_s\to  \pi^{+} \rho^{-}$, $B_s\to  \pi^{-} \rho^{+}$, $B_s\to  \pi^{0} \rho^{0}$, $B_s\to  \pi^{0} \omega$ and $B_s\to  \pi^{0} \pi^{0}$.

\subsection{CP asymmetry study}

\begin{table}
\caption{The direct CP asymmetries ($\mathcal{A}$) and
 mixing-induced CP asymmetries   ($\mathcal{S}$)
    of $\bar B\to PP$ decays. We also show the results from conventional flavor diagram approach \cite{Cheng:2014rfa}  for comparison. }\label{tab:CPBPP}
\begin{center}
\scriptsize\begin{tabular}{cccccccc}
\hline\hline
  & Mode &  ~~~~~$\mathcal{A_{\rm exp}}$~~ & ~~$\mathcal{A}_{this~work}$~~~~~~&$\mathcal{A}_{Flavor~diagram}$~~~~&
~~~~~~~~$\mathcal{S_{\rm exp}}$~~&~~~$\mathcal{S}_{this~work}$~~&~~$\mathcal{S}_{Flavor~diagram}$~~\\
\hline
&$ \pi^{+} \pi^{-}$     &$\star0.31\pm0.05$   &$0.31\pm0.04$     &  $0.326\pm0.081$    &$\star-0.67\pm0.06$&   $-0.60\pm0.03$   &    $-0.717\pm0.061$       \\
&$  \pi^{0}  \pi^{0}$    &$0.43\pm0.24$         &$0.57\pm0.06$        &  $0.611\pm0.113$      &$     $&        $0.58\pm0.06$  &      $0.454\pm0.112$       \\
&$  \pi^{0} \eta$        &$     $               &$-0.16\pm0.16$       &  $0.566\pm0.114$      &$     $&        $-0.98\pm0.04$&       $-0.098\pm0.338$       \\
&$  \pi^{0} \eta^{'}$    &$     $               &$0.39\pm0.14$        &  $0.385\pm0.114$      &$     $&        $-0.90\pm0.07$ &      $0.142\pm0.234$       \\
&$ \eta \eta$            &$     $               &$-0.85\pm0.06$       &   $-0.405\pm0.129$   &$     $&         $0.33\pm0.12$ &       $-0.796\pm0.077$       \\
&$ \eta \eta^{'}$        &$     $               &$-0.97\pm0.04$       &   $-0.394\pm0.117$   &$     $&         $-0.20\pm0.15$ &      $-0.903\pm0.049$       \\
&$ \eta^{'} \eta^{'}$    &$     $               &$-0.87\pm0.07$       &   $-0.122\pm0.136$   &$     $&         $-0.46\pm0.14$ &      $-0.964\pm0.037$       \\
&$  \pi^{0}  K_s$ &$0.00\pm0.13$     &$-0.14\pm0.03$    &  $-0.173\pm0.019$    &$\star0.58\pm0.17$ &    $0.73\pm0.01$  &      $0.754\pm0.014$         \\
&$ \eta  K_s$     &$     $               &$-0.30\pm0.10$       &  $-0.301\pm0.041$  &$     $         &  $0.68\pm0.04$  &      $0.592\pm0.035$         \\
&$ \eta^{'}  K_s$ &$0.06\pm0.04$     &$0.030\pm0.004$   &  $0.022\pm0.006$   &$\star 0.63\pm0.06$   &   $0.69\pm0.00$  &      $0.685\pm0.004$          \\
&$ K^{0} \bar{K^{0}}$    &                      &$-0.057\pm0.002$     & $0.017\pm0.041$     &$0.8\pm0.5$     & $0.099\pm0.002$   &         0      \\
\hline
&$ \pi^{-}\pi^{0}$  &$0.03\pm0.04$    &$-0.026\pm0.003$     &  $0.069\pm0.027$      &$     $&           $ $   &            \\
&$ \pi^{-} \eta$         &$-0.14\pm0.07$                     &$-0.14\pm0.07$       &  $-0.081\pm0.074$     &$     $&           $ $   &           \\
&$ \pi^{-} \eta^{'}$     &$0.06\pm0.16$                      &$0.37\pm0.07$        &  $0.374\pm0.087$      &$     $&           $ $   &           \\
&$ \pi^{-} \bar{K^{0}}$  &$-0.017\pm0.016$                   &$0.0027\pm0.0001$        &   0                   &$     $         &$ $   &           \\
&$  \pi^{0} K^{-}$       &$0.037\pm0.021$                    &$0.065\pm0.024$          &   $0.047\pm0.025$    &$     $         &$ $   &           \\
&$ \eta K^{-}$           &$\star -0.37\pm0.08$               &$-0.22\pm0.08$     &   $-0.426\pm0.043$   &$     $         &$ $   &           \\
&$ \eta^{'} K^{-}$       &$0.013\pm0.017$                    &$-0.021\pm0.007$         &   $-0.027\pm0.008$   &$     $         &$ $   &           \\
&$ K^{-} K^0$          &$-0.21\pm0.14$                       &$-0.057\pm0.002$        &       0               &$     $         &$$   &           \\
&$ \pi^{+} K^{-}$        &$\star -0.082\pm0.006$             &$-0.081\pm0.005$   &   $-0.080\pm0.011$   &$     $         &$$   &           \\
\hline\hline
\end{tabular}
\end{center}
\end{table}

The charmless B decays are important mostly because of its large direct CP asymmetry in B decays.
Due to the CKM matrix elements suppression of tree diagram, the penguin diagram contribution is at the same order magnitude as the tree diagram. The large CKM phase difference between these two kinds of diagram almost guarantees the existence of  large direct CP asymmetry. That is not the whole story.  The direct CP asymmetry parameter is also proportional to the strong phase difference between these two diagrams. Unfortunately, the strong phase is mostly from non-perturbative QCD dynamics. That is the reason why the QCD factorization and soft-collinear effective theory can predict the branching ratios of the charmless B decays well but make wrong prediction or no prediction for the direct CP asymmetries.
There are already 3 good measurements of direct CP asymmetry measurements in $B\to PP$ decays and 3 in $B\to PV$ decays indicated as a star in Tables~\ref{tab:CPBPP} and \ref{tab:CPBudPV}.  There are also 5 mixing induced CP asymmetry measurements for the neutral B meson decays to be used in our $\chi^2$ program.
 We give the direct $CP$
and mixing-induced $CP$ asymmetries of corresponding B decay modes in
Tables~\ref{tab:CPBPP} and \ref{tab:CPBudPV}.
From the CP asymmetry formula in eq.(\ref{cp}), we know that the CP asymmetry is proportional to the difference of $B$ meson and $\bar B$ meson. Thus the theoretical uncertainty from hadronic parameters mostly cancel,  because they contribute to the charge conjugate modes equally. The main theoretical uncertainty for CP asymmetry parameters is from the experimental data and CKM angle.
We did not show the individual uncertainty, but the combined one in these CP asymmetry tables.

\begin{table}
\caption{The direct CP asymmetries ($\mathcal{A}$) and
 mixing-induced CP asymmetries   ($\mathcal{S}$)
    of $\bar B\to PV$ decays. We also show the results from conventional flavor diagram approach \cite{Cheng:2014rfa}  for comparison. }\label{tab:CPBPV}
\label{tab:CPBudPV}
\begin{center}
\scriptsize\begin{tabular}{cccccccc}
\hline\hline
  & Mode &  ~~~~~$\mathcal{A_{\rm exp}}$~~ & ~~$\mathcal{A}_{this~work}$~~~~~~&~~$\mathcal{A}_{Flavor~diagram}$~~~~&
~~~~~~~~$\mathcal{S_{\rm exp}}$~~&~~~$\mathcal{S}_{this~work}$~~&~~$\mathcal{S}_{Flavor~diagram}$~~~~\\
\hline
&$ \pi^{+} \rho^{-}$        &$0.13\pm0.06$              &$0.15\pm0.03$          &  $0.120\pm0.027$              &$  0.07\pm0.14 $ &$0.011\pm0.034$      &  $-0.049\pm0.074$         \\
&$ \pi^{-} \rho^{+}$        &$-0.08\pm0.08$             &$-0.44\pm0.03$         &  $-0.136\pm0.053$             &$  0.05\pm0.08 $ &$-0.093\pm 0.040$    &  $-0.024\pm0.065$         \\
&$  \pi^{0} \rho^{0}$       &$-0.27\pm0.24$             &$0.36\pm0.08$          &  $-0.043\pm0.121$             &$-0.23\pm0.34$   &$0.19\pm0.16$      &  $-0.229\pm0.112$         \\
&$  \pi^{0} \omega$         &$     $                    &$-0.024\pm0.068$       &  $-0.188\pm0.185$             &$     $          &$0.29\pm0.05$      &  $-0.315\pm0.195$             \\
&$ \eta \rho^{0}$           &$     $                    &$-0.23\pm0.03$         &  $-0.264\pm0.215$             &$     $          &$-0.023\pm0.038$  &  $-0.628\pm0.196$         \\
&$ \eta \omega$             &$     $                    &$-0.30\pm0.13$         &  $0.054\pm0.137$              &$     $          &$0.43\pm0.09$      &  $-0.461\pm0.113$         \\
&$ \eta^{'} \rho^{0}$       &$     $                    &$0.088\pm0.085$        &  $-0.440\pm0.317$             &$     $          &$-0.48\pm0.07$     &  $-0.714\pm0.252$         \\
&$ \eta^{'} \omega$         &$     $                    &$-0.85\pm0.17$         &  $-0.005\pm0.259$             &$     $          &$0.50\pm0.26$      &  $-0.624\pm0.120$         \\
&$ K_s \rho^{0}$     &$0.04\pm0.20$      &$-0.085\pm0.059$        &   $0.069\pm0.053$             &$0.5\pm0.21$           &$0.88\pm0.05$          &    $0.643\pm0.036$       \\
&$  K_s \omega$       &$0\pm0.4$          &$0.25\pm0.10$           &   $-0.053\pm0.055$            &$\star 0.7\pm0.21$     &$0.70\pm0.04$    &    $0.789\pm0.028$       \\
&$  K_s \phi$         &$-0.01\pm0.14$     &$-0.006\pm0.001$      &    0                          &$\star 0.59\pm0.14$      &$0.70\pm0.00$       &    $0.718\pm0.000$       \\
&$ \bar{K^{0}} K^{*0}$       &$     $            &$-0.10\pm0.02$          &    0                          &$     $                &$ -0.90\pm0.03 $      & 0\\
&$ K^{0} \bar{K^{*0}}$       &$     $            &$-0.18\pm0.01$          &    0                          &$     $                &$0.89\pm0.03 $       & 0 \\
\hline
&$ \pi^{-} \rho^{0}$        &$0.18^{+0.09}_{-0.17}$     &$-0.45\pm0.04$         &  $-0.239\pm0.084$             &$     $          &$ $                &            \\
&$ \pi^{-} \omega$          &$-0.04\pm0.06$             &$0.054\pm0.052$      &  $0.075\pm0.067$              &$     $          &$$                 &           \\
&$  \pi^{0} \rho^{-}$       &$0.02\pm0.11$              &$0.16\pm0.02$          &  $0.053\pm0.094$              &$     $          &$$                 &           \\
&$ \eta \rho^{-}$           &$0.11\pm0.11$              &$-0.11\pm0.02$         &  $0.162\pm0.072$              &$     $          &$$                 &           \\
&$ \eta^{'} \rho^{-}$       &$0.26\pm0.17$              &$0.45\pm0.05$          &  $0.223\pm0.137$              &$     $          &$$                 &           \\
&$ \pi^{-} \bar{K^{*0}}$     &$-0.04\pm0.09$     &$0.005\pm0.001$       &    0                          &$     $             &$$          &              \\
&$  \pi^{0} K^{*-}$          &$-0.06\pm0.24$     &$0.088\pm0.040$         &   $-0.116\pm0.092$            &$     $             &$$          &           \\
&$ \eta K^{*-}$              &$0.02\pm0.06$      &$-0.17\pm0.02$          &   $-0.016\pm0.037$            &$     $             &$$          &           \\
&$ \eta^{'} K^{*-}$          &$-0.26\pm0.27$     &$-0.45\pm0.09$          &   $-0.391\pm0.162$            &$     $             &$$          &           \\
&$ K^{-} \rho^{0}$           &$\star0.37\pm0.10$      &$0.59\pm0.06$      &   $0.306\pm0.100$             &$     $             &$$          &           \\
&$ K^{-} \omega$             &$0.02\pm0.05$      &$0.19\pm0.09$           &   $0.010\pm0.080$             &$     $             &$$          &           \\
&$ K^{-} \phi$               &$0.04\pm0.04$      &$-0.006\pm0.001$      &    0                          &$     $             &$$          &           \\
&$ K^{-} K^{*0}$             &$     $            &$-0.10\pm0.02$          &     0                          &$     $             &$$          &           \\
&$ K^0 K^{*-}$             &$     $            &$-0.18\pm0.01$          &      0                         &$     $             &$$          &           \\
&$ \bar{K^0} \rho^{-}$     &$-0.12\pm0.17$     &$0.009\pm0.000$       &    0                          &$     $             &$$          &           \\
&$ \pi^{+} K^{*-}$           &$\star-0.22\pm0.06$     &$-0.20\pm0.04$     &   $-0.217\pm0.048$            &$     $             &$$          &           \\
&$  \pi^{0} \bar{K^{*0}}$    &$-0.15\pm0.13$     &$-0.27\pm0.05$          &   $-0.332\pm0.114$            &$     $             &$$          &           \\
&$ \eta \bar{K^{*0}}$        &$\star0.19\pm0.05$      &$0.065\pm0.011$    &   $0.099\pm0.028$             &$     $             &$$          &           \\
&$ \eta^{'} \bar{K^{*0}}$    &$-0.07\pm0.18$     &$0.059\pm0.049$         &   $0.069\pm0.152$             &$     $             &$$          &           \\
&$ K^{-} \rho^{+}$           &$0.21\pm0.11$      &$0.59\pm0.01$           &   $0.134\pm0.053$             &$     $             &$$          &           \\
\hline\hline
\end{tabular}
\end{center}
\end{table}

Since the CKM matrix elements are enhanced for penguin diagram compared with
  the tree diagrams in the $B \to \pi(\rho) K^{(*)}$ decays  by $b\to s$ transition, there is large interference effect between these two kinds of Feynman diagrams, which results in larger CP asymmetry in these decays.
$A_{CP}$($\bar B^{0}\to \pi^{+} {K^{-}}$) is the first measurement of direct CP asymmetry in B decays.
 From table \ref{tab:BrBudPP}, one can see that $ B^{-} \to \pi^{0} {K^{-}}$ decay has the same dominant decay amplitude $T$ and $P$ as $\bar B^{0}\to \pi^{+} {K^{-}}$ decay, thus one expects the same direct CP asymmetry \cite{Cheng:2009cn}.   However, experimentally these two direct CP asymmetry is quite different, even with an opposite sign. That is the so-called $\pi K$ CP-puzzle. In our study, the sub-leading contribution $C$ and $P_{EW}$ are not negligible, especially $C$ with a large strong phase, therefore this puzzle
is   resolved.

There is one category of decays with pure penguin contributions, such as $B^-\to K^-K^0$, $\bar B^0\to K^0 \bar K^0$, $B^-\to \pi^-\bar K^0$, $B^-\to \pi^- \bar K^{*0}$, $B^-\to \rho^-\bar K^0$ and $\bar B_s \to K^0\bar K^0$. Their direct CP asymmetry is expected to be zero, at leading order approximation. The very small (not zero) CP asymmetry is from the small up quark or charm quark penguin contribution interference with the dominant top quark contribution. Any large CP asymmetry measurement for these decays will be a clear signal of new physics.
In Table \ref{tab:CPBPP}, we did not  show the decay channel  $\bar B^0 \to K^+K^-$. The reason is that there should be  two major contributions for this channel, but we calculate only one (tree level W exchange digram). The other contribution from penguin-exchange ($P_E$) diagram is not fitted because of lack of experimental data. The branching ratio of this channel with only one contribution,  discussed in previous subsection,   is far from the central value of experimental data. This may indicate the importance of the penguin-exchange ($P_E$) diagram, which will give a large direct CP asymmetry for this channel. Similarly, we can not predict the CP asymmetry for $B^0 \to K^{*+}K^-$ and
$B^0 \to K^{+}K^{*-}$.

\begin{table}
\caption{The direct CP asymmetries ($\mathcal{A}$) and
 mixing-induced CP asymmetries   ($\mathcal{S}$)
    of $\bar B_s\to PP$ decays. We also show the results from conventional flavor diagram approach \cite{Cheng:2014rfa}  for comparison. }\label{tab:CPBsPP}
\begin{center}
\scriptsize\begin{tabular}{cccccccc}
\hline\hline
  & Mode &  ~~~~~$\mathcal{A_{\rm exp}}$~~ & ~~$\mathcal{A}_{this~work}$~~~~~~&$\mathcal{A}_{Flavor~diagram}$~~~~&
~~~~~~~~$\mathcal{S_{\rm exp}}$~~&~~~$\mathcal{S}_{this~work}$~~&~~$\mathcal{S}_{Flavor~diagram}$~~\\
\hline
&$  \pi^{0} \eta$           &$  $         &$0.90\pm0.05$     & $-0.165\pm0.292$   &$     $    &$0.19\pm0.11$            &   $0.836\pm0.198$        \\
&$  \pi^{0} \eta^{'}$       &$     $      &$0.44\pm0.10$     & $0.259\pm0.335$    &$     $    &$-0.79\pm0.07$           &   $0.953\pm0.116$        \\
&$  \pi^{0}  K_s$          &$     $      &$0.87\pm0.05$    &  $0.724\pm0.054$    &$     $    &$0.0096\pm0.0905$        & $0.302\pm0.080$          \\
&$ \eta \eta$               &$     $      &$-0.11\pm0.01$    & $-0.116\pm0.018$   &$     $    &$-0.14\pm0.01$           &   $-0.095\pm0.020$        \\
&$ \eta \eta^{'}$           &$     $      &$-0.013\pm0.005$  & $-0.009\pm0.003$   &$     $    &$-0.038\pm0.006$         &   $-0.036\pm0.007$        \\
&$ \eta K_s$              &$     $        &$0.74\pm0.17$     & $0.452\pm0.057$    &$     $    &$0.31\pm0.16$            & $0.787\pm0.042$          \\
&$ \eta^{'} \eta^{'}$       &$     $      &$0.042\pm0.006$   & $0.016\pm0.009$    &$     $    &$-0.055\pm0.006$         &   $0.028\pm0.009$        \\
&$ \eta^{'}  K_s$          &$     $        &$-0.58\pm0.06$    & $-0.367\pm0.089$   &$     $    &$-0.029\pm0.099$         & $0.191\pm0.090$          \\
&$ K^{+} K^{-}$             &$-0.14\pm0.11$   &$-0.11\pm0.02$   & $-0.090\pm0.021$  &$0.30\pm0.13$   &$0.097\pm0.022$    &   $0.140\pm0.030$        \\
&$ K^{0} \bar{K^{0}}$       &$     $         &$0.0027\pm0.0001$ & $-0.075\pm0.035$  &$     $         &$0.069\pm0.000$    &   $-0.039\pm0.001$        \\
\hline
&$ \pi^{-} K^{+}$           &$0.28\pm0.04$                   &$0.16\pm0.01$      &    $0.266\pm0.033$     &$     $        &$$                  &           \\
\hline\hline
\end{tabular}
\end{center}
\end{table}

The mixing induced $CP$ asymmetries in neutral $B$ decays into final $CP$
eigenstates are dominated by the $B^0 -\bar B^0$ mixing phase with little dependence on strong phases. That is the reason why it is usually used for searching possible new physics.
For example, the measured mixing induced CP asymmetry parameters of $S_{CP}(\pi^{+}\pi^{-})$, $S_{CP}(\pi^{0} K_{S})$, $S_{CP}(\eta^{\prime}K_{S})$ and
$S_{CP}(\phi K_{S})$ have received much attention in experiment and in theoretical aspect
due to little   theoretical uncertainty. Currently, there is a good agreement between theoretical calculations and experimental data shown in Table \ref{tab:CPBPP} and \ref{tab:CPBPV}. Further study is needed from
both theoretical  and   experimental effort in the future.

\begin{table}
\caption{The direct CP asymmetries ($\mathcal{A}$) and
 mixing-induced CP asymmetries   ($\mathcal{S}$)
    of $\bar B_s\to PV$ decays. We also show the results from conventional flavor diagram approach \cite{Cheng:2014rfa}  for comparison. }\label{tab:CPBsPPPV}
\label{tab:CPBsPV}
\begin{center}
\footnotesize\begin{tabular}{cccccc}
\hline\hline
  & Mode &  ~~~~~~~$\mathcal{A}_{this~work}$~~~~~~&~~$\mathcal{A}_{Flavor~diagram}$~~~~&
~~~~~~$\mathcal{S}_{this~work}$~~&~~$\mathcal{S}_{Flavor~diagram}$~~~~\\
\hline
&$  \pi^{0} \phi$             &$0.89\pm0.04$         &   $0.073\pm0.201$             &$-0.25\pm0.07$    &   $0.439\pm0.171$      \\
&$ \eta \rho^{0}$             &$-0.46\pm0.38$        &   $0.323\pm0.136$           &$0.88\pm0.19$     &   $-0.002\pm0.168$      \\
&$ \eta \omega$              &$-0.086\pm0.071$       &   $-0.432\pm0.271$           &$-0.31\pm0.06$    &   $-0.238\pm0.296$      \\
&$ \eta \phi$                 &$0.083\pm0.113$        &   $0.428\pm0.504$           &$0.39\pm0.15$     &   $0.534\pm0.400$      \\
&$ \eta^{'} \rho^{0}$         &$-0.67\pm0.10$        &   $0.323\pm0.136$            &$-0.72\pm0.07$    &   $-0.002\pm0.168$      \\
&$ \eta^{'} \omega$           &$0.33\pm0.06$         &   $-0.432\pm0.271$            &$-0.14\pm0.07$    &   $-0.238\pm0.296$      \\
&$ \eta^{'} \phi$             &$-0.010\pm0.017$    &   $0.043\pm0.090$             &$0.047\pm0.015$   &   $0.166\pm0.057$      \\
&$ K^{+} K^{*-}$            &$-0.30\pm0.04$        &   $-0.217\pm0.048$            &$  -0.78\pm0.06$                & 0        \\
&$ K^{-} K^{*+}$           &$0.39\pm0.04$        &   $0.134\pm0.053$              &$0.67\pm0.05$                &  0       \\
&$ K_s \rho^{0}$            &$-0.42\pm0.15$        &   $-0.124\pm0.453$          &$0.78\pm0.08$     &   $-0.348\pm0.285$      \\
&$ K_s \omega$             &$-0.010\pm0.151$      &   $-0.029\pm0.436$              &$-0.32\pm0.30$  &   $0.928\pm0.110$      \\
&$ K_s \phi$             &$-0.003\pm0.033$       &     0                          &$-0.85\pm0.01$    &   $-0.692\pm0.000$      \\
&$ K^0 \bar{K^{*0}}$       &$0.002\pm0.001$     &     0                         &$ -0.74\pm0.05 $                &   0      \\
&$ \bar{K^0} K^{*0}$      &$0.009\pm0.000$     &     0                          &$     0.83\pm0.04$                &   0        \\
\hline
&$ \pi^{-} K^{*+}$        &$-0.30\pm0.01$        &   $-0.136\pm0.053$             &$$                &         \\
&$  \pi^{0} K^{*0}$       &$-0.30\pm0.06$        &   $-0.423\pm0.158$            &$$                &         \\
&$ \eta K^{*0}$            &$0.57\pm0.12$         &   $0.828\pm0.123$            &$$                &         \\
&$ \eta^{'} K^{*0}$        &$-0.46\pm0.10$       &   $-0.408\pm0.273$           &$$                &         \\
&$ K^{+} \rho^{-}$          &$0.16\pm0.03$        &   $0.120\pm0.027$             &$ $               &         \\
\hline\hline
\end{tabular}
\end{center}
\end{table}

There are only two channels of $B_s$ decays, namely $B_s\to K^+K^-$ and $\bar B_s\to K^+\pi^-$ with CP asymmetry measurements shown in table \ref{tab:CPBsPP}. As stated, we do not include any $B_s$ data in our $\chi^2$ fit. All the $B_s$ results are predictions. It is easy to see that our predictions for these two channels agree with data within error-bar. There is no CP asymmetry measurement for $B_s \to PV$ decays. Our theoretical predictions are shown in table \ref{tab:CPBsPV}, together with results from the conventional flavor diagram approach. It is noted that there is large differences between  predictions of these two approaches for example: $B_s\to  \pi^0 \phi$, $B_s\to \eta \rho^0$, $B_s\to \eta' \rho^0$ and $B_s\to \eta '\omega$ etc. Many of these entries with large CP asymmetry predicted, can be tested by the experiments in the near future. Similar to the situation of branching ratios, we also did not give predictions for the CP asymmetry of decays $B_s\to  \pi^{+} \pi^{-}$, $B_s\to  \pi^{+} \rho^{-}$, $B_s\to  \pi^{-} \rho^{+}$, $B_s\to  \pi^{0} \rho^{0}$, $B_s\to  \pi^{0} \omega$ and $B_s\to  \pi^{0} \pi^{0}$, lack of the information of penguin exchange diagram ($P_E$).

\subsection{The flavor SU(3) asymmetry}

The flavor $SU(3)$ symmetry is broken by the difference
in the u, d and s quark masses, especially the difference in
d and s quark masses. The $SU(3)$ breaking is also very important in explaining the  different size of $CP$ asymmetry in different charmless $B\to PP$, $PV$ decays.
We consider the flavor $SU(3)$ violating contributions assisted
by factorization hypothesis where the source of $SU(3)$ asymmetries
are mainly from decay constants and weak transition form factors.
It is not necessary to include different $SU(3)$ asymmetry phases
for different modes, because our numerical results of branching ratios and CP asymmetry parameters are in good agreement with experimental
data     shown in previous subsections.

As every decay mode include various topological diagrams,
the precise flavor $SU(3)$ breaking effect, can not be
separated from one another in $B\to PP, PV$ decays,
are hard to be tested by experimental data.
We show the flavor $SU(3)$ breaking effect in every topology
amplitude between $B \to \pi\pi $ and $B \to \pi K $, $B \to \eta\pi $ and $B \to \eta K $,
  as following:
\begin{align}
 \label{SU(3)T}
 \vert\frac{{T}(B^{-}\to \pi^{0}K^{-})}{V_{ub}V_{us}^{*}}\vert:\vert\frac{T (B^{-}\to \pi^{0}\pi^{-})}{V_{ub}V_{ud}^{*}}\vert =1:0.83 ,\\
 \label{SU(3)C}
 \vert\frac{C(B^{-}\to \pi^{0}K^{-})}{V_{ub}V_{us}^{*}}\vert:\vert\frac{C(B^{-}\to \pi^{0}\pi^{-})}{V_{ub}V_{ud}^{*}}\vert=1:0.91 ,\\
 \label{SU(3)P}
 \vert\frac{P(\bar{B^{0}}\to \pi^{+}K^{-})}{V_{tb}V_{ts}^{*}}\vert:\vert\frac{P(\bar{B^{0}}\to \pi^{+}\pi^{-})}{V_{tb}V_{td}^{*}}\vert=1:0.89 ,\\
 \label{SU(3)PC}
 \vert\frac{P_C(B^{-}\to \eta K^{-})}{V_{tb}V_{ts}^{*}}\vert:\vert\frac{P_C(B^{-}\to \eta \pi^{-} )}{V_{tb}V_{td}^{*}}\vert  =1:0.91 .
\end{align}
From the above results, we find that the flavor $SU(3)$ breaking effects
are   around $10\%$ because of different
decay constants between $f_{\pi}$ and $f_{K}$,  or form factors
$F^{B\to \pi}$ and $F^{B\to K}$. The flavor $SU(3)$ breaking effect in every topology
amplitude between
$B \to \pi\rho $ and $B \to \pi K^{*}$,  $B \to \eta \rho $ and $B \to \eta K^{*}$ are also shown as following:
\begin{align}
  \label{SU(3)Tv}
  \vert\frac{T(B^{-}\to \pi^{0}K^{*-})}{V_{ub}V_{us}^{*}}\vert:\vert\frac{T(B^{-}\to \pi^{0}\rho^{-})}{V_{ub}V_{ud}^{*}}\vert=1:0.83,\\
  \label{SU(3)Cv}
  \vert\frac{C(B^{-}\to K^{*-}\pi^{0})}{V_{ub}V_{us}^{*}}\vert:\vert\frac{C(B^{-}\to \rho^{-}\pi^{0})}{V_{ub}V_{ud}^{*}}\vert=1:0.80 ,  \\
  \label{SU(3)Pv}
  \vert\frac{P(\bar{B^{0}}\to \pi^{+}K^{*-})}{V_{tb}V_{ts}^{*}}\vert:\vert\frac{P(\bar B^{0}\to \pi^{+}\rho^{-})}{V_{tb}V_{td}^{*}}\vert=1:0.74, \\
  \label{SU(3)PCv}
  \vert\frac{P_C(B^{-}\to K^{*-} \eta )}{V_{tb}V_{ts}^{*}}\vert:\vert\frac{P_C (B^{-}\to \rho^{-}\eta  )}{V_{tb}V_{td}^{*}}\vert=1:0.80 ,\\
  \label{SU(3)PAv}
  \vert\frac{P_A(\bar{B^{0}}\to \pi^{+}K^{*-})}{V_{tb}V_{ts}^{*}}\vert:\vert\frac{P_A(\bar B^{0}\to \pi^{+}\rho^{-})}{V_{tb}V_{td}^{*}}\vert=1:0.84.
\end{align}
It is easy to see that the flavor $SU(3)$ breaking effects are larger  than $20\%$ because of different
decay constants between $f_{\rho}$ and $f_{K^*}$,  or different form factors between
$A_0^{B\to \rho}$ and $A_0^{B\to K^*}$.

In previous flavor diagram approach, the charmless $B\to PP$ and $B\to PV$ decays are fitted separately with very different theoretical parameters. That implies large difference between pseudo-scalar meson and vector meson.
To show this difference numerically, we have:
 \begin{align}
\label{SU(3)Tpirho} \vert T (B^{-}\to \pi^{0}\pi^{-})\vert:
 \vert T (B^{-}\to \pi^{0}\rho^{-})\vert  =1 : 1.64  ,\\
\label{SU(3)Cpirho} \vert C (B^{-}\to \pi^{0}\pi^{-})\vert:
 \vert C^\prime (B^{-}\to \pi^{-}\rho^{0})\vert =1: 1.43  ,\\
\label{SU(3)Ppirho}\vert P (\bar{B^{0}}\to \pi^{+}\pi^{-})\vert:
 \vert P (\bar{B^{0}}\to \pi^{+}\rho^{-})\vert =1: 0.66 .
\end{align}
It is easy to see that this difference between $\pi$ and $\rho$ meson emission is indeed much larger than the so called flavor SU(3) breaking effect between $\pi$ and $K$ meson, because the meson decay constant $f_\rho > f_K$.  The penguin amplitude (P) for the $\bar B\to \pi^+ \rho^-$ decay, even if with a larger decay constant, is smaller than the corresponding $\bar B\to \pi^+\pi^-$ decay, because there is no chiral enhanced penguin contribution for a vector meson emission shown in eq.(\ref{eq:P}).
If the emitted meson is a pseudo-scalar scalar meson in $B \to VP$ decays, its difference from $B\to PP$ decays is the $B\to V$ transition form factor from $B\to P$ transition form factor. For example, the following difference between two decay channels is $B\to \pi$ form factor and $B\to \rho$ form factor, which is smaller than the difference between $\pi$ and $\rho$ decay constant:
 \begin{align}
\label{SU(3)Tpirho2} \vert T (B^{-}\to \pi^{0}\pi^{-})\vert:
   \vert T (B^{-}\to \rho^{-} \pi^{0})\vert =1  : 1.24,\\
\label{SU(3)Cpirho2} \vert C (B^{-}\to \pi^{0}\pi^{-})\vert:
  \vert C (B^{-}\to \rho^{-} \pi^{0})\vert =1  : 1.25,\\
\label{SU(3)Ppirho2}\vert P (\bar{B^{0}}\to \pi^{+}\pi^{-})\vert:
 \vert P (\bar B^{0}\to \rho^{-} \pi^{+})\vert =1 : 0.59,\\
\label{SU(3)PCpirho} \vert P_C (B^{-}\to \eta \pi^{-})\vert:
 \vert P_C(B^{-}\to \rho^{-}\eta )\vert=1: 1.26.
\end{align}
The penguin amplitude (P) for the $\bar B\to  \rho^-\pi^+$ decay, even if with a larger decay constant, is smaller than the corresponding $\bar B\to \pi^+\pi^-$ decay, because the   chiral enhanced penguin contribution cancel some of the  factorization penguin contribution as a minus sign  shown in eq.(\ref{eq:P}).
For decays induced by $b\to s$ transition, we have:
\begin{align}
\label{SU(3)TKKstar} \vert T (B^{-}\to \pi^{0} K^{-})\vert:
 \vert T (B^{-}\to \pi^{0} K^{*-})\vert =1 : 1.42,\\
 \label{SU(3)CKKstar}\vert C (B^{-}\to \pi^{0} K^{-})\vert:
 \vert C (B^{-}\to K^{*-}\pi^{0} )\vert =1 : 1.23,\\
\label{SU(3)PKKstar}\vert P (\bar{B^{0}}\to \pi^{+} K^{-})\vert:
 \vert P (\bar{B^{0}}\to \pi^{+} K^{*-})\vert =1: 0.68,\\
\label{SU(3)PCKKstar} \vert P_C (B^{-}\to \eta K^{-})\vert:
 \vert P_C(B^{-}\to K^{*-}\eta )\vert=1: 1.24.
\end{align}
It is apparent that the difference characterized by the $ K $ and $ K^*$  decay constant is large.

\section{Conclusion}\label{conclusion}

In this paper, we studied two-body charmless hadronic B decays in factorization assisted
topological amplitude approach. Since factorization has been proven to all orders in $\alpha_{s}$ in the so called soft-collinear effective theory
at leading order in $\Lambda/m_b$ expansion, the color-favored tree
emission diagram $T$ was factorized into short-distance
effective Wilson coefficients and decay constants and form factors, without free parameters.
The flavor $SU(3)$ breaking effects are then automatically considered in different meson decay constants and transition form factors.
 Factorization theorem is not proven in most other topological diagrams. They   were considered as universal magnitudes ($\chi$)
and associated phases ($\phi$) in the conventional flavor diagram approach to be fitted from experimental data.  In our approach, the corresponding decay constants, form factors were
factorized out from them before $\chi^2$ fit assisted by factorization hypothesis to indicate
the flavor $SU(3)$ breaking effect.
In addition to the large tree and QCD-penguin
  diagrams studied in these types of decays, the electro-weak
penguin contribution ($P_{EW}$) was also included, which is not negligible
but essential for $B \to \pi K$ decays, especially for the CP asymmetry parameters.   Unlike the previous conventional flavor diagram approach, this contribution was factorized into short-distance
effective Wilson coefficients and decay constants and form factors, just like the color-favored tree
emission diagram $T$.

There were 6 parameters $\chi^{C}(\phi^{C}),\chi^{C^{\prime}}(\phi^{C^{\prime}})$
and $\chi^{E}(\phi^{E})$ for tree diagrams $C,E$ and 8 parameters $\chi^{P}(\phi^{P}),
\chi^{P_C}(\phi^{P_C}), \chi^{P_C^{\prime}}(\phi^{P_C^{\prime}})$ and $\chi^{P_A}(\phi^{P_A})$
for QCD-penguin diagrams to be fitted from 48 measured data of branching ratios and CP asymmetry parameters. Since $SU(3)$ breaking effects and the difference between pseudo-scalar and vector meson have been already considered in the decays constants and form factors, we can fit all the $B \to PP$, $PV$ decays together. The number of free parameters is greatly reduced.
These parameters
were extracted precisely even for small parameters $\chi^{E},\phi^{E}$, which had
large uncertainties in conventional flavor diagram approach. Besides, the $\chi^{2}$
per degree of freedom  is smaller than the conventional flavor diagram approach, even with much more free parameters in their approach.
With the fitted parameters, we predicted   branching fractions of $B_{(s)} \to PP$,  $PV$
decay modes and their $CP$ asymmetry parameters. The long-standing puzzles of $\pi\pi$ branching ratios and $\pi K$ $CP$ asymmetry  have been resolved consistently
with not too large color suppressed tree diagram contribution $\chi^{C}$. 
For the $B_{s}$ decays, we do not include any data as input in the $\chi^2$ fit, but all as theoretical predictions, since very  few channels have been poorly measured. The flavor $SU(3)$ breaking effect
between $\pi$ and $K$ were approximately $10\%$, even more than $20\%$
in $\rho$ and $K^{*}$ meson case and larger in $\pi$ and $\rho$, $K$ and $K^*$.

\section{Acknowledgments}
We are grateful to Hsiang-nan Li, Wei Wang, Rui Zhou, Fusheng Yu, Ying Li and Qin Qin for useful discussion.
We also thanks Fred James providing help for analyzing error bar in Minuit program.
The work is partly supported by National   Science Foundation of China (11375208, 11521505, 11621131001 and 11235005).


\end{document}